\documentclass[final,3p,times]{elsarticle}
\usepackage{graphicx}
\usepackage{booktabs}
\usepackage{cite}
\usepackage{picinpar}
\usepackage{amsmath,amsfonts,amssymb,amsthm}
\usepackage{orcidlink}
\usepackage{url}
\usepackage{subcaption}
\usepackage[linesnumbered,ruled,vlined]{algorithm2e}

\usepackage{algpseudocode}

\let\oldnl\nl
\newcommand{\nonl}{\renewcommand{\nl}{\let\nl\oldnl}}
\usepackage{bm}
\usepackage{soul}
\usepackage{colortbl}
\usepackage{multirow, makecell}
\usepackage{comment}
\usepackage{pifont}
\usepackage{tikz}
\usepackage{pgfplots}
\usetikzlibrary{shapes, arrows, positioning, arrows.meta, calc, backgrounds, fit}
\pgfplotsset{compat=1.18}
\usepackage{color}
\usepackage{alltt}
\usepackage{hyperref}

\hypersetup{colorlinks,citecolor=blue,colorlinks=true,linkcolor=blue, urlcolor=blue, linktocpage}
\usepackage{enumerate}
\usepackage{siunitx}
\usepackage{hyperref}
\usepackage{epstopdf}
\usepackage{pbox}
\newtheorem{problem}{Problem}
\newtheorem{theorem}{Theorem}

\newtheorem{lemma}{Lemma}
\newtheorem{remark}{Remark}
\newtheorem{assumption}{Assumption}

\usepackage{framed} 
\usepackage{multicol} 
\biboptions{sort&compress}

\journal{Engineering Applications of Artificial Intelligence}
\begin{document}

\begin{frontmatter}

\title{Physics-informed sparse identification-based tube model predictive control for aerial vehicles}

 \author[label1]{Tayyab Manzoor\orcidlink{0000-0003-3932-0506}}
 \author[label2,label3]{Yasir Ali\orcidlink{0000-0003-3998-9218}}
 \author[label4,label2]{Yuanqing Xia\orcidlink{0000-0002-5977-4911}\corref{mycorrespondingauthor}}\ead{xia\_yuanqing@bit.edu.cn} 
\cortext[mycorrespondingauthor]{Corresponding Author}
\author[label1,label2]{Lijie You\orcidlink{0000-0003-0724-8861}}
 \author[label1]{Yan Wang\orcidlink{0000-0001-9070-6653}}

\address[label1]{School of Automation and Electrical Engineering, Zhongyuan University of Technology, Zhengzhou, Henan 450007, China}
\address[label2]{School of Automation, Beijing Institute of Technology, Beijing 100081, China}
 \address[label3]{Zhengzhou Academy of Intelligent Technology, Beijing Institute of Technology, Zhengzhou 450000, China}
 \address[label4]{Zhongyuan University of Technology, Zhengzhou, Henan 450007, China}
\begin{abstract}
Autonomous aerial vehicles necessitate control strategies that balance computational efficiency with robust performance in dynamic operational environments. This paper proposes a model predictive control (MPC) framework for aerial platforms that leverages physics-informed machine learning (PIML) to achieve an optimal balance between computational tractability and robust performance. At the core of the proposed approach lies a sparse, control-affine model identified via the PIML method, which provides a parsimonious yet interpretable representation of the system dynamics by embedding first-principles knowledge and learning residual uncertainties from operational data. This model is incorporated within a robust MPC scheme that adopts a high-order Runge-Kutta discretization to ensure prediction accuracy and an adaptive tube-based mechanism to guarantee constraint satisfaction under uncertainty. The online adaptation of the tube, directly informed by the residual error of the PIML model, ensures robust stability without introducing excessive conservatism. Rigorous theoretical proofs are provided to establish recursive feasibility and stability. Numerical simulations and experiments on a quadrotor demonstrate that our method significantly reduces computational load compared to nonlinear MPC and robust MPC using a first-principles high-fidelity model, while outperforming PID, nonlinear MPC, neural-network-based MPC, and fixed-tube robust MPC in tracking performance and robustness, showcasing the practical efficiency of the proposed PIML-based control synthesis for resource-constrained aerial systems.
\end{abstract}

\end{frontmatter}
\begin{table*}[!htbp]   
\begin{framed}
\centering
\small
\textbf{Nomenclature}
\vspace{4pt} 
\begin{multicols}{2}
\raggedright
\noindent
$\bullet^\mathcal{B}, \bullet^\mathcal{I}$ \quad Body-fixed and inertial frame\\
$\hat{\bm \bullet}, \bm{\bullet}^*$ \quad Learned and optimal value of $\bm{\bullet}$\\
$\mathbb{N}, \mathbb{R}$ \quad Set of natural and real numbers\\
$\|(\bm{\bullet})\|$ \quad Euclidean norm\\
$\|(\bm{\bullet})\|_{Q}$ \quad Q-weighted matrix norm\\
$\bm{\bullet}\left(j | k\right)$ \quad $\bm\bullet$ at $j$ predicted at time $k$\\
$\oplus, \ominus$ \quad Minkowski sum, Pontryagin difference\\
$s_\bullet, c_\bullet, t_\bullet$ \quad $\sin(\bullet), \cos(\bullet), \tan(\bullet)$\\
$\mathcal{D}, \mathbb{D}$ \quad Learning dataset, disturbance set\\
$\mathcal{U}_{\xi}$ \quad Learning-induced uncertainty set
\end{multicols}
\end{framed}
\end{table*}
\section{Introduction}
Aerial vehicles are emerging as transformative technologies, revolutionizing transportation, industrial automation, and smart urban development. Their applications in autonomous logistics, precision agriculture, and infrastructure inspection are driving significant advancements, reshaping industries, and enabling innovative solutions to complex challenges \citep{XU2026111789, TIAN2026111085, CHENG2026111513}. However, despite their numerous advantages, the widespread adoption of drones presents several challenges. While widely used PID controllers are effective for simple, single-objective tasks like stabilizing an aerial vehicle's altitude or position, they lack the predictive, adaptive, and collaborative capabilities required for complex tasks \citep{OSMAN2026111646}. Despite their potential, Machine Learning (ML) techniques alone are insufficient because they often lack the predictive capabilities and real-time adaptability required for dynamic environments, as their purely data-driven nature lacks the fundamental physical constraints and first-principles reasoning necessary for reliable prediction and adaptation. Model predictive control (MPC), on the other hand, helps to mitigate several issues associated with non-optimal control methods \citep{YE2026111790}.

However, MPC law is intrinsically tied to the accuracy of the underlying dynamic model for model-based optimization. Traditionally, controllers for aerial vehicles are developed using first-principles models derived from physics-informed models (PIMs) \citep{drones7010004}. While physically intuitive, these nominal models are often insufficiently precise, and they fail to capture complex, hard-to-model effects. This model mismatch manifests as persistent disturbances that can degrade tracking performance, violate safety constraints, and, in worst-case scenarios, lead to system instability. The field of ML offers powerful tools for learning residual dynamics from operational data by embedding PIM, leading to the paradigm of Physics-Informed Machine Learning (PIML) \citep{safe_learning}. Several works have successfully demonstrated the use of PIML involving neural network (NN)-based models for model identification \citep{NN-mpc1, NN-MPC3, 11124830, hong2023physics, 9834096, 10770871, drones9030187, ARYAN2026114379}. Nevertheless, NN-based methods require large training data and are generally not interpretable \citep{tayyab_tnnls1, controlsystem_letter}. Furthermore, another challenge remains in how to certify robust constraint satisfaction and stability for an MPC controller based on a learned model that inherently possesses residual uncertainty. This challenge can be addressed by implementing hybrid control algorithms that leverage the capabilities of ML and MPC techniques \citep{active_learning, 10606056, TIE2}, enabling more robust, adaptive, and precise drone operations in complex environments. To address the issue associated with NN-based control approaches, the sparse identification of nonlinear dynamics (SINDy) driven MPC control technique can be adopted \citep{drones7010004, https://doi.org/10.1002/rnc.70272, tayyab_tnnls1, IET_control, ICCTA}. While promising, generally, SINDy techniques suffer due to the manual library selection. Moreover, real-world adoption of PIML techniques for aerial vehicles faces critical challenges in real-time implementation, generalization, data scarcity, interpretability, and computational efficiency. Robust MPC (RMPC) frameworks are designed to handle bounded uncertainties \citep{rmpc1, Robust_mpc2}. A common approach is tube-based MPC \citep{tube_1, tube_2, tube_3, tube_4}, which uses a nominal prediction model and a feedback controller to confine the actual state within a robust positively invariant (RPI) tube around the nominal trajectory. The key to this method is the accurate characterization of the disturbance set, which bounds the discrepancy between the nominal model and the true system. Most existing RMPC approaches assume that the disturbance set is known a priori and constant. This is a significant limitation when using a learned model, as the model error is not constant; it is initially large, decreases as more data is collected, and may vary with the operating regime. A non-adaptive, overly conservative disturbance bound would lead to poor performance, while an optimistic one would jeopardize robustness.

\begin{figure}[ht!]
    \centering
    \includegraphics[width=.85\linewidth]{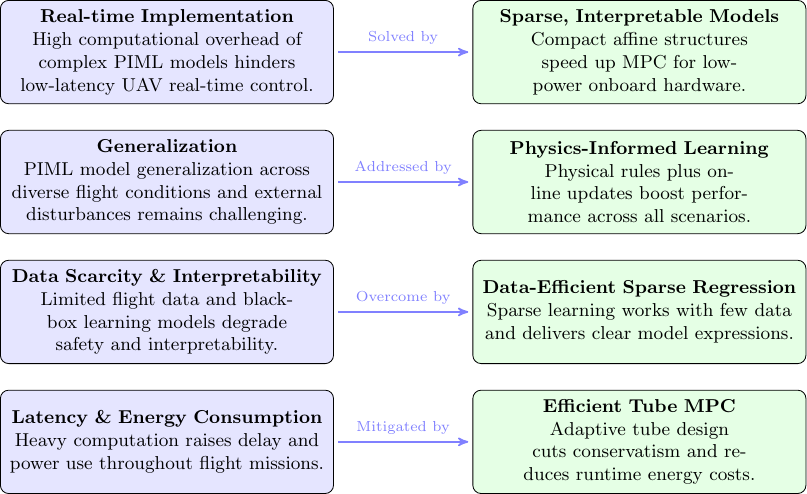}
    \caption{Mapping of key challenges in PIML-based MPC for aerial vehicles to our proposed solutions. Each challenge (left) is directly addressed by a specific contribution (right), demonstrating how our framework overcomes practical deployment
barriers.}
    \label{fig:fig01}
\end{figure}

To bridge this gap, this paper proposes a PIML-based tube MPC framework for aerial vehicles that addresses key challenges for PIML-based MPC for aerial vehicles, shown in Fig. \ref{fig:fig01}.
Our core contribution is a methodology that seamlessly leverages online model learning with robust control, providing formal guarantees for systems in which the model is refined and its uncertainty quantified in real time. As summarized in Table \ref{tab:method_summary}, all existing SINDy-based MPC frameworks for aerial vehicles rely on manually predefined function libraries and do not integrate any tube-based robust constraint mechanism. Different from these works, the proposed method realizes fully automatic library construction and sparse pruning for SINDy identification. More importantly, we combine the sparse learning model with a real-time adaptive tube scheme, which is absent in current SINDy-MPC studies. In addition, we provide complete theoretical guarantees for the overall learning-control system and conduct both numerical simulations and real hardware experiments, further distinguishing our work from existing approaches. 
\begin{table*}[h]
\centering
\caption{Comparison of existing SINDy-based MPC methods and the proposed approach for aerial vehicles}
\footnotesize
\begin{tabular}{ll}
\hline
\textbf{Method}& \begin{tabular}[c]{@{}l@{}}\textbf{Description (Model Learning, Tube Adaptation,}  \textbf{Theoretical}\\ \textbf{Proof, Real-time Implementation)}\end{tabular} \\
\hline
RMPC \citep{drones7010004} & \begin{tabular}[c]{@{}l@{}}Adaptive SINDy with fixed library; no tube structure; stability and\\ feasibility proofs; simulation-only validation. \end{tabular}\\
\hline
RMPC \citep{tayyab_tnnls1} &\begin{tabular}[c]{@{}l@{}} Offline SINDy combined with RL (static library); no tube structure;\\ stability and feasibility proofs; simulation-only validation. \end{tabular}\\
\hline
Lyapunov-based MPC \citep{https://doi.org/10.1002/rnc.70272} & \begin{tabular}[c]{@{}l@{}}Adaptive SINDy for coefficient update (fixed library); no tube \\structure; Lyapunov stability proof; validated via both simulation\\ and real-time experiments. \end{tabular}\\
\hline
Conventional MPC \citep{IET_control} & \begin{tabular}[c]{@{}l@{}}Static physics-guided SINDy library with offline identification; no\\ tube structure; only model accuracy analysis (no proofs);\\ simulation-only validation.\end{tabular} \\
\hline
Standard MPC \citep{ICCTA} & \begin{tabular}[c]{@{}l@{}}Fixed manual SINDy library without automatic pruning; no tube\\ structure; no theoretical proofs; validated only via simulation.\end{tabular} \\
\hline
Proposed Method & \begin{tabular}[c]{@{}l@{}}Physics-informed automatic library + online sparse pruning; adaptive\\ tube mechanism; complete stability and feasibility proofs; validated\\ via both simulation and real-time experiments.\end{tabular} \\
\hline
\end{tabular}
\label{tab:method_summary}
\end{table*}
In other words, the contributions of this work are threefold:
\subsection{Contributions}
\begin{enumerate}
    \item Unlike NN-based PIML approaches \citep{NN-mpc1, NN-MPC3, 11124830, hong2023physics, 9834096, 10770871, drones9030187, ARYAN2026114379}, that require large datasets and lack interpretability, we employ sparse regression to learn a compact, physically consistent residual model. This balances accuracy with real-time computational feasibility. Contrary to traditional SINDy-based MPC methods that rely on manually designed function libraries and fixed dictionary structures \citep{drones7010004, https://doi.org/10.1002/rnc.70272, tayyab_tnnls1, IET_control, ICCTA}, the proposed framework realizes fully automatic library initialization, nonlinear expansion, and sparse pruning based on quadrotor rigid-body mechanics. It eliminates artificial intervention in library construction and improves model adaptability while retaining the interpretability and data efficiency of sparse identification. This fundamentally differs from both conventional manually tuned SINDy-MPC and pure data-driven learning-based MPC. 
    \item Many existing tube-based MPCs adopt constant or empirically preset disturbance sets, leading to excessive conservatism or insufficient robustness. In this work, the disturbance bound is recursively updated using real-time model residual errors and external disturbance measurements, so the tube size adapts to time-varying model uncertainty and wind turbulence. This design breaks the limitation of static bounds in traditional robust tube MPC.
    \item While many PIML-based MPC methods lack stability proofs under online adaptation \citep{IET_control, ICCTA}, we provide rigorous recursive feasibility and Input-to-State Stability (ISS) guarantees for the closed-loop system, ensuring safety and stability throughout the learning process. We also provide numerical simulations and experimental results, as opposed to \citep{drones7010004,tayyab_tnnls1, IET_control, ICCTA}, to corroborate the efficacy of the developed approach.  
    \item Different from existing SINDy-MPC with high computational overhead and NN-MPC with large memory consumption, the periodic sparse regression and low-complexity RPI set update ensure the proposed method can run stably on low-power STM32 microcontrollers, achieving a good trade-off among tracking accuracy, robustness, and computational cost.
\end{enumerate}

The remainder of this manuscript is organized as follows. Section \ref{section2} formulates the problem with details of the PIML dynamics identification process and control methodology. Section \ref{section3} is dedicated to the theoretical analysis, proving recursive feasibility and stability. Section \ref{section4} presents numerical simulation results and experiments demonstrating the efficacy of the proposed approach. Finally, Section \ref{section5} provides a summary of key findings and suggests avenues for future research.
\section{Problem formulation and control methodology}\label{section2}
This section establishes the mathematical foundation for our approach, beginning with the PIML dynamics model and proceeding to the MPC formulation.
\subsection{PIML dynamics}
The derivation of the PIML dynamics model for the aerial vehicle proceeds in two sequential steps. First, the PIM is established, and this foundational step distills the system's dynamics into a structured analytical framework. As defined in Fig.~\ref{fig:0.1}, two coordinate frames are introduced: the body-fixed frame $\mathcal{B}\triangleq\{{o^{\mathcal{B}}}, x^{\mathcal{B}}, y^{\mathcal{B}}, z^{\mathcal{B}}\}$ and the inertial frame $\mathcal{I}\triangleq\{{o^{\mathcal{I}}}, x^{\mathcal{I}}, y^{\mathcal{I}}, z^{\mathcal{I}}\}$. The vehicle's dynamics are governed by a set of differential equations derived from rigid-body mechanics. The PIM encapsulates this first-principles model, defining the state variables, control inputs, and the fundamental structure of the dynamic relationships. This model provides the architectural prior for the subsequent ML component and is provided as follows:
\begin{figure}[ht!]
    \centering
    \includegraphics[width=.6\linewidth]{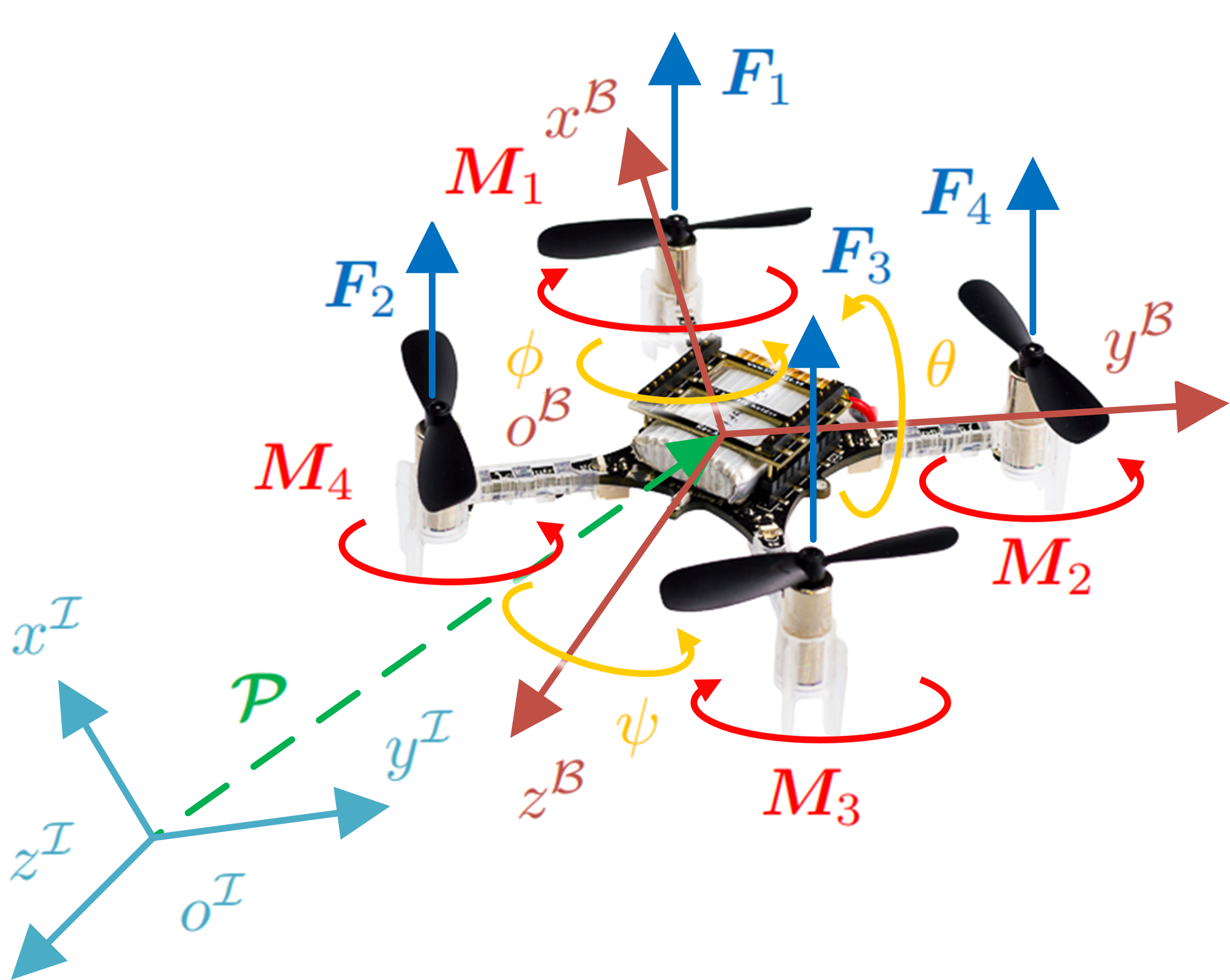}
    \caption{Aerial vehicle's inertial and body-fixed frame}
    \label{fig:0.1}
\end{figure}
\begin{equation}\label{eq01}
\begin{aligned}
    \dot{\mathcal{\bm{P}}}&=[\dot{\mathcal{P}}_x~ \dot{\mathcal{P}}_y~ \dot{\mathcal{P}}_z]^\top=\bm{v}^{\mathcal{I}}=[v_{x}^{\mathcal{I}}~v_{y}^{\mathcal{I}}~v_{z}^{\mathcal{I}}]^\top, \\
    \dot{\bm{v}^{\mathcal{I}}}&=\frac{1}{m}(\underbrace{[0 ~0~ mg]^{\top}}_{\text{Gravity}}+\bm{R}^{\mathcal{BI}}(\Theta)\underbrace{[0 ~0~ u_1]^\top}_{\text{Thrust}}),\\
    \dot{\bm{\Theta}}&=\begin{bmatrix}
        \dot{\phi}\\
        \dot{\theta}\\
        \dot{\psi}
    \end{bmatrix}=\underbrace{\begin{bmatrix}
        1 &s_\phi t_{\theta} &c_{\phi}t_{\theta} \\
         0&c_{\phi} &-s_{\phi}\\
         0&s_{\phi}/c_{\theta}&c_{\phi}/c_{\theta}
    \end{bmatrix}}_{\text{Euler angle rate matrix}}\bm{\omega}^{\mathcal{B}},\\
    \dot{\bm{\omega}}^{\mathcal{B}}&=\bm{J}^{-1}(\underbrace{[u_2~u_3~u_4]^\top}_{\text{Torque}}-\bm{\omega}^{\mathcal{B}}\times \bm{J}\bm{\omega}^{\mathcal{B}}),
    \end{aligned}
\end{equation}
with
$$u_1=\sum_{i=1}^{4}F_i,~ \bm{J}=\text{diag}\begin{bmatrix}
  J_{xx}~J_{yy}~ J_{zz}  
\end{bmatrix}^\top,$$
\begin{equation}\notag
\begin{aligned}
&\begin{bmatrix}
    u_2\\
    u_3\\
    u_4
\end{bmatrix}=\begin{bmatrix}
    l(F_2-F_4)\\
    l(F_3-F_1)\\
    \kappa(F_1-F_2+F_3-F_4)
\end{bmatrix},\\ &\bm{R}^{\mathcal{BI}}(\Theta)=\begin{bmatrix}
  c_{\theta} c_{\psi}& c_{\theta} s_{\psi}& -s_{\theta} \\
   s_{\phi}s_{\theta}c_{\psi}-c_{\phi}s_{\psi}& s_{\phi}s_{\theta}c_{\psi}+c_{\phi}c_{\psi}&s_\phi c_\theta\\
    c_{\phi}s_{\theta}c_{\psi}+s_{\phi}s_{\psi}& c_{\phi}s_{\theta}s_{\psi}-s_{\phi}c_{\psi}& c_{\phi}c_{\theta}   
\end{bmatrix}, 
\end{aligned}
\end{equation}
where $\bm{\mathcal{P}}$, $\bm{v}^{\mathcal{I}}$, $\bm{\omega}^{\mathcal{B}}$, $m$, $\bm{\Theta}$ are the position, linear velocity, angular velocity, mass, and attitude of the aerial vehicle, respectively. $\phi$, $\theta$, $\psi$ are the Euler angles. Moreover, $s_\bullet = \sin(\bullet)$, $c_\bullet = \cos(\bullet)$, $t_\bullet = \tan(\bullet)$. $l$ and $\kappa$ are arm length and Torque-to-thrust ratio, respectively. $\bm{\Delta}_v$ and $\bm{\Delta}_\omega$ are the learned parts in the translational and rotational dynamics of the aerial vehicle, respectively. Any deviation in the vehicle dynamics is handled by the ML approach, that is utilized to learn the time-series data, where data is collected, which consists of measurements of the state $\bm{x}=[\bm{\mathcal{P}}$, $\bm{\omega}^{\mathcal{B}}$,  $\bm{v}^{\mathcal{I}}$, $\bm{\Theta}]^\top$ and input $\bm{u}=[u_1, u_2, u_3, u_4]^\top$, which includes evaluating the required number of snapshots $n_s$ of $\bm{x}$, $\bm{u}$. Hence, the dataset $\mathcal{D}$ can be described in the following two matrices $\hat{\bm{x}}$ and $\hat{\bm{u}}$. 
\begin{equation}\label{x_s}
\begin{aligned}
\hat{\bm{x}}=&\left[\begin{array}{cccc}
\mathcal{P}^{\mathcal{I}}_x\left(t_1\right) & \mathcal{P}^{\mathcal{I}}_y\left(t_1\right) & \cdots & \psi\left(t_1\right) \\
\vdots & \vdots & \ddots & \vdots \\
\mathcal{P}^{\mathcal{I}}_x\left(t_i\right) & \mathcal{P}^{\mathcal{I}}_y\left(t_i\right) & \cdots & \psi\left(t_i\right) \\
\vdots & \vdots & \ddots & \vdots \\
\mathcal{P}^{\mathcal{I}}_x\left(t_{n_s}\right) & \mathcal{P}^{\mathcal{I}}_y\left(t_{n_s}\right) & \cdots & \psi\left(t_{n_s}\right)
\end{array}\right],\\
\hat{\bm{u}}=&  \left[\begin{array}{cccc}
u_{1}\left(t_1\right) & u_{2}\left(t_1\right) & u_{3}\left(t_1\right) & u_{4}\left(t_1\right) \\
\vdots & \vdots & \vdots & \vdots \\
u_{1}\left(t_i\right) & u_{2}\left(t_i\right) & u_{3}\left(t_i\right) & u_{4}\left(t_i\right) \\
\vdots & \vdots & \ddots & \vdots \\
u_{1}\left(t_{n_s}\right) & u_{2}\left(t_{n_s}\right) & u_{3}\left(t_{n_s}\right) & u_{4}\left(t_{n_s}\right)
\end{array}\right], 
\end{aligned}
\end{equation}
where $t_i$ represents the sampling time. Therefore, the numerical differentiation yields the quadrotor's state as follows: 
\begin{equation}\label{x_d}
\dot{\hat{\bm{x}}}=\left[\begin{array}{cccc}
\dot{\mathcal{P}}^{\mathcal{I}}_x\left(t_1\right) & \dot{\mathcal{P}}^{\mathcal{I}}_y\left(t_1\right) & \cdots & \dot{\psi}\left(t_1\right) \\
\vdots & \vdots & \ddots & \vdots \\
\dot{\mathcal{P}}^{\mathcal{I}}_x\left(t_i\right) & \dot{\mathcal{P}}^{\mathcal{I}}_y\left(t_i\right) & \cdots & \dot{\psi}\left(t_i\right) \\
\vdots & \vdots & \ddots & \vdots \\
\dot{\mathcal{P}}^{\mathcal{I}}_x\left(t_{n_s}\right) & \dot{\mathcal{P}}^{\mathcal{I}}_y\left(t_{n_s}\right) & \cdots & \dot{\psi}\left(t_{n_s}\right)
\end{array}\right].
\end{equation}
Thereafter, a library function $\bm{\Psi}(\bm{\hat{x}}, \bm{\hat{u}})$ is constructed using physics-informed rigid-body dynamic terms, including $\hat{\bm{x}}$ and $\hat{\bm{u}}$, their first-order products, and pairwise multiplicative interactions. No manual customization or domain-specific heuristic terms are introduced, preserving fully automated and interpretable sparse modeling for quadrotor dynamics, which is provided as follows:
\begin{equation}\label{5sindy3}
\begin{aligned}
\bm{\Psi}(\bm{\hat{x}}, \bm{\hat{u}})=[\bm{1}^\top, \bm{\hat{x}}^\top, (\bm{\hat{x}} \otimes \bm{\hat{x}})^\top, (\bm{\hat{x}} \otimes \bm{\hat{u}})^\top, \cdots, s_{\bm{\hat{x}}^\top},  s_{\bm{\hat{u}}^\top}, s_{\bm{\hat{x}} \otimes \bm{\hat{u}}^\top}, \cdots],
\end{aligned}
\end{equation}
where $\otimes$ represents the product combination of the components. Therefore, the sparse model can be expressed as follows:
\begin{equation}\label{5sindy4}
    \hat{f}(\bm{\hat{x}}, \bm{\hat{u}}, S_j)=\sum_{k=0}^{n_{c_f}}S_{\ell j}\delta_{\ell}(\bm{\hat{x}}, \bm{\hat{u}}), ~ j=1,2,3,4, \dots , 12, 
\end{equation}
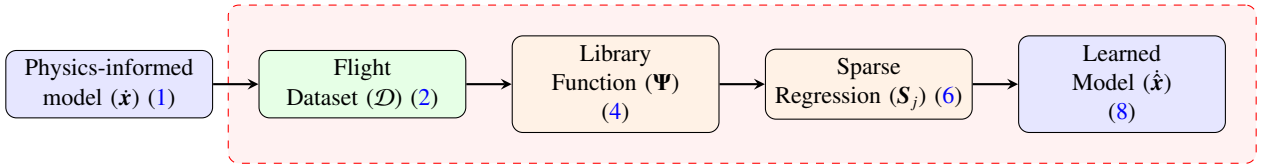
\begin{figure*}[!htbp]
    \centering
    \makebox[\textwidth][c]{
\begin{tikzpicture}[
    >=stealth,
    block/.style={rectangle, draw, fill=blue!10, text width=2.5cm, align=center, minimum height=.8cm, rounded corners, font=\small},
    data/.style={rectangle, draw, fill=green!10, text width=2.5cm, align=center, minimum height=0.8cm, rounded corners, font=\small},
    process/.style={rectangle, draw, fill=orange!10, text width=2.5cm, align=center, minimum height=.8cm, rounded corners, font=\small},
    arrow/.style={->, thick},
    node distance=1.2cm and 0.6cm 
]

\node (UAV) [block] {Physics-informed\\model ($\dot{\bm{x}}$) \eqref{eq01}};
\node (DataAcq) [data, right=of UAV] {Flight\\ Dataset ($\mathcal{D}$) \eqref{x_s}};
\node (Library) [process, right=of DataAcq] {Library\\ Function ($\bm{\Psi}$)\\\eqref{5sindy3}};
\node (SINDy) [process, right=of Library] {Sparse\\ Regression ($\bm{S}_j$) \eqref{5sindy5}};
\node (Model) [block, right=of SINDy] {Learned\\ Model ($\dot{\hat{\bm{x}}}$)\\\eqref{5nominal_system}};

\draw [arrow] (UAV.east) -- (DataAcq.west);
\draw [arrow] (DataAcq.east) -- (Library.west);
\draw [arrow] (Library.east) -- (SINDy.west);
\draw [arrow] (SINDy.east) -- (Model.west);

\begin{scope}[on background layer]
    \node [draw=red, dashed, fill=red!5, inner sep=0.4cm, rounded corners,
           fit=(DataAcq) (Library) (SINDy) (Model)] (IdentificationProcess) {};
\end{scope}

\end{tikzpicture}
}
 \caption{PIML framework}
    \label{fig:3.0}
\end{figure*}
where $n_{c_f}$, $S_{\ell k}$, $\delta_\ell$ denote the number of candidate functions, candidate functions of $\ell$-th column in $\bm{\Psi}(\bm{\hat{x}}, \bm{\hat{u}})$ and weighted coefficient of $\delta_{\ell}(\bm{\hat{x}}, \bm{\hat{u}})$ concerned with $j$-th state, i.e., $\bm{S}_j=[S_{j0}, S_{j1}, S_{j2}, \dots, S_{jn_{c_f}}]^\top$, respectively. Next, sparse optimization is used as follows:
\begin{equation}\label{5sindy5}
\bm{S}_j=\underset{\bm{S}_j}{\arg \min }\left\|\dot{\bm{\hat{x}}}_{j}-\bm{S}_j \bm{\Psi}(\bm{\hat{x}}, \bm{\hat{u}})^\top\right\|_2+h_j\left\|\bm{S}_j\right\|_1,
\end{equation}
where $\dot{\bm{\hat{x}}}_{j}$, $h_j$ are the $j$-th row of $\dot{\bm{\hat{x}}}$ represent hyper-parameter, respectively. $\bm{S}_k$ is stored in matrix $\bm{\xi}$, 
\begin{equation}
   \bm{\xi} =\left[\bm{S}_1 \bm{S}_2 \cdots \bm{S}_j \cdots \bm{S}_{12}\right].
\end{equation}
Therefore, the learned model can be defined as:
\begin{equation}\label{5nominal_system}
    \dot{\hat{\bm{{x}}}}=\hat{f}(\bm{\hat{x}}, \bm{\hat{u}})\approx\bm{\Psi}(\bm{\hat{x}}, \bm{\hat{u}})\bm{\xi},
\end{equation}
where $\hat{f}(\bm{\hat{x}}, \bm{\hat{u}})$ is the learned model. The complete process is depicted in Fig.~\ref{fig:3.0}. The library function selection is an automatic process. The initial library is populated with universal rigid-body mechanics terms (e.g., $\hat{\bm{x}}$, $\hat{\bm{u}}$, etc.) and is not handpicked but fundamental to modeling quadrotor motion (e.g., position, velocity, control input interactions). Nonlinear candidate functions are auto-added when the flight dataset $\mathcal{D}$ exceeds $N_{min}=500$ samples. $N_{min}$ is empirically tuned via preliminary flight tests to balance noise filtering and computational overhead for auto library expansion. This trigger ensures the library adapts to unmodeled dynamics (e.g., ground effect, wind) without manual intervention, and no user decides which terms to include, so the automatic property of the library is fully preserved. The relevant terms are discarded automatically via regression, leaving only terms that contribute to the dynamic modeling. This step prevents library bloat and ensures sparsity, with no manual curation of useful functions. Furthermore, flight data is preprocessed via a complementary filter (IMU + barometer) to reduce sensor noise, followed by 3 times outlier rejection, which is standard practice for UAV state estimation in order to ensure data quality for sparse regression. Outliers exceeding 3 times the data standard deviation are discarded to avoid biasing the sparse model coefficients.
\begin{remark}
The efficacy of sparse identification is indeed contingent on the candidate functions in the regressor library, but this does not imply manual selection. In standard SINDy, the base library is manually defined (e.g., polynomials up to degree 3, trigonometric terms), introducing subjectivity and risk of incomplete/poorly chosen sets. Our framework eliminates this limitation via a fully automatic library construction pipeline. 
\end{remark}
It is worth clarifying that preloading universal rigid-body base terms originates from inherent first-principle physical laws of quadrotor dynamics instead of empirical manual selection for specific flight scenarios, which differs fundamentally from conventional SINDy requiring hand-designed domain-specific library dictionaries. While physical base formulas act as fixed universal prior knowledge, all additional cross-coupling and nonlinear candidate functions are automatically supplemented when the dataset size exceeds $N_{\mathrm{min}}$, and redundant terms are autonomously eliminated through sparse regression without manual curation, validating the fully automatic library screening property of our framework. To further elaborate on the fully automatic library construction rules for reproducibility, the details are given as follows:
\begin{itemize}
    \item Initial library: The base candidate library is composed of universal rigid-body dynamic terms, including system states, control inputs, and their Kronecker products, which are determined by inherent quadrotor mechanics without manual selection.
    \item Nonlinear expansion rule: Trigonometric and high-order cross-coupling terms are automatically generated following fixed rules once the dataset size exceeds the preset threshold $N_{\mathrm{min}}=500$.
    \item Threshold $N_{\mathrm{min}}$: This value is calibrated via preliminary flight tests to balance noise suppression and library scale and is fixed for all formal experiments. 
    \item Sparse Regularization parameter: A unified regularization coefficient is obtained via cross-validation on preliminary flight data and remains constant during all trials.
    \item Pruning criteria: After sparse regression, candidate terms with absolute coefficients smaller than the preset cutoff $10^{-4}$ are automatically pruned as redundant. The entire workflow requires no manual intervention, which distinguishes our design from traditional SINDy methods with manually customized libraries.
\end{itemize}
The real continuous-time system of the vehicle is expressed as follows:
\begin{equation}
    \dot{\bm{x}}=f(\bm{x}, \bm{u})+\bm{\Delta}_{vw}(t),
\end{equation}
where $\bm{\Delta}_{vw}(t)\in \mathbb{W}$ is a bounded external factor, which may involve external wind turbulence, unmodelled uncertainties, and noise \citep{tayyab_tnnls1}. Also, $\bm x(t) \in \mathbb{X} \subset \mathbb{R}^{12}$ and $\bm u(t) \in \mathbb{U} \subset \mathbb{R}^4$. The sets $\mathbb{X}$ and $\mathbb{U}$ are convex and compact, and $\mathbb{W}$ is compact and contains the origin.
\\\textbf{Control objective}: Design a feedback control law that minimizes the infinite-horizon cost $\sum_{0}^{\infty} \ell_{c}(\bm x, \bm u)$ while ensuring all closed-loop trajectories satisfy the state and input constraints $x_k \in \mathbb{X}$, $\bm u_k \in \mathbb{U}$ for all $k$, despite the disturbance. The stage cost $\ell_c(\bm x, \bm u)$ is typically quadratic: $\ell_c(\bm x, \bm u)=(\bm x-\bm x_{\text{ref}})^\top\bm Q(\bm x-\bm x_{\text{ref}})+(\bm u-\bm u_{\text{ref}})^\top\bm R(\bm u-\bm u_{\text{ref}})$ with $\bm Q \succeq 0$, $\bm R \succ 0$.
\begin{remark}
The core challenge in PIML-based control is maintaining robust constraint satisfaction despite time-varying model uncertainty that increases as more data are collected. Existing robust MPC approaches typically assume constant disturbance bounds, which leads to either excessive conservatism or insufficient robustness when applied to learning-based models.
\end{remark}
\begin{assumption}
The PIML model uncertainty can be decomposed as $\dot{\hat{\bm{x}}} = \bm{\Psi}(\hat{\bm{x}}, \hat{\bm{u}})(\bm{\xi} + \Delta\bm{\xi}) + \bm{\Delta}_{vw}(t)$, where $\Delta\bm{\xi}$ represents parametric uncertainty in the learned coefficients and $\bm{\Delta}_{vw}(t)$ captures external disturbances. Both are bounded with known conservative bounds initially. This is realistic for aerial vehicles where model inaccuracies arise from both imperfect identification (parametric) and environmental factors (external).  
\end{assumption}
The discrete nominal dynamics after using the Runge-Kutta (RK4) method can be defined as follows:
\begin{equation}
    \hat{\bm{x}}(k+1)=\hat{f}^d(\hat{\bm{x}}(k), \hat{\bm{u}}(k))+\bm{\Delta}^d_{vw}(k),
\end{equation}
where $\bm{\Delta}^d_{vw}(k) \in \mathbb{D}(k)$ captures the rest of the uncertainties, including in the PIML model, RK4 discretization, or any external disturbances. The set $\mathbb{D}(k)$ is assumed to be compact and contain the origin. In the second step, compute the realized disturbance sample (model error + external wind + sensor noise):
\begin{equation}\label{step2_adaption}
    \bm{\Delta}^d_{vw}(k)=\bm x(k+1)-\hat{f}^d(\bm{x}(k), \bm{u}(k)). 
\end{equation}
Next, we define the error between nominal and actual states $\tilde{\bm{e}}(k)=\bm{x}(k)-\hat{\bm{x}}(k)$. Applying the controller $\tilde{\bm{u}}(k)=\hat{\bm{x}}(k)+K\tilde{\bm{e}}(k)$, the error dynamics can be defined as:
\begin{equation}\label{eq12error}
\begin{aligned}
&\tilde{\bm{e}}(k+1)\approx \tilde{\bm A}(k)\tilde{\bm{e}}(k)+\bm{\Delta}^d_{vw}(k)+\bm{\Delta}_{\xi}(k)
\end{aligned}
\end{equation}
where $\tilde{\bm A}(k)= \frac{\partial \hat{f}^d}{\partial \hat{\bm{x}}} \big|_{(\hat{\bm{x}}(k), \hat{\bm{u}}(k))}$ is the Jacobian of the discrete-time model. $\bm{\Delta}_{\xi}(k)\in\mathcal{U}_{\xi}(k)$ is the learning-induced uncertainty from PIML parameter updates $\Delta\bm \xi(k)=\bm \xi(k+1)-\bm \xi(k)$. The set $\mathcal{U}_{\xi}(k)$ is defined as:
\begin{equation}\mathcal{U}_{\xi}(k)=\mathcal{L}_{\xi}.||\Delta\bm \xi(k)||.\mathbb{B}_r \subset \mathbb{R}^{12},
\end{equation}
where $\mathcal{L}_{\xi}>0$ is the Lipschitz constant of the discrete PIML model $\hat{f}^d$ with respect to the learned coefficients $\bm \xi$ (exists by smoothness of rigid-body quadrotor dynamics and sparse regression), and $\mathbb{B}_r={z\in \mathbb{R}^{12}:||z||\leq 1}$ is the unit Euclidean ball. A set $\mathbb{S}\subset \mathbb{R}^{12}$ is a time-varying RPI set for the augmented error dynamics if: \begin{equation}\label{eq14R}
\tilde{\bm A}(k)\mathbb{S}(k)\oplus\mathbb{D}(k)\oplus\mathcal{U}_{\xi}(k) \subseteq\mathbb{S}(k).
\end{equation}
This guarantees $\tilde{\bm{e}}(k) \in \mathbb{S}(k) \implies \tilde{\bm{e}}(k+1) \in \mathbb{S}(k+1)$ for all $k\geq0$ no monotonicity required-only boundedness. \\
Tightened Constraints: To ensure robust constraint satisfaction, Tightened State Constraints: $\mathbb{X}_\mathbb{S}(k) = \mathbb{X} \ominus \mathbb{S}(k)$. Tightened Input Constraints: $\mathbb{U}_\mathbb{S}(k) = \mathbb{U} \ominus K\mathbb{S}(k)$, where $K$ is the tube feedback gain.
The disturbance set $\mathbb{D}(k)$ is a hypercube centered at $c(k) \in \mathbb{R}^{12}$ with element-wise bounds $\bar{\Delta}_{vw}^d(k) \in \mathbb{R}_{>0}^{12}$:
\[
\mathbb{D}(k) = \left\{ \Delta \in \mathbb{R}^{12} : |\Delta_i - c_i(k)| \leq \bar{\Delta}_{vw,i}^d(k), \forall i=1,...,12 \right\}.
\]
To ensure $\mathbb{D}(k)$ is compact and noise-robust, we update the center and bounds recursively:

Step 1: Update Disturbance Set Center (Exponential Smoothing)
\begin{equation}
c(k+1) = \lambda c(k) + (1-\lambda) \bm\Delta_{vw}^d(k), 
\end{equation}
where $\lambda \in (0,1)$ (tuned to $\lambda=0.9$ for the aerial vehicle) is an exponential smoothing gain that filters high-frequency sensor noise.

Step 2: Update Disturbance Bounds (Forgetting Factor + Clipping)
\begin{equation}
\bar{\bm \Delta}_{vw}^d(k+1) = \gamma \bar{\bm \Delta}_{vw}^d(k) + (1-\gamma) \left\| \bm \Delta_{vw}^d(k) - c(k+1) \right\|_\infty, \end{equation}
\begin{equation}
\bar{\bm \Delta}_{vw}^d(k+1) = \bar{\Delta}_{max} \quad \text{if} \quad \bar{\bm \Delta}_{vw}^d(k+1) > \bar{\bm \Delta}_{max}, 
\end{equation}
where $\gamma \in (0,1)$ (tuned to $\gamma=0.95$) is a forgetting factor that replaces the non-decreasing max operator (fixes unbounded growth), $\|\cdot\|_\infty$ is the infinity norm (robust to outlier disturbance samples), $\bar{\bm \Delta}_{max} \in \mathbb{R}_{>0}^{12}$ (set to $\bar{\bm \Delta}_{max,i}=0.1$ for all $i$) is a hard upper bound derived from quadrotor physics (max wind/noise/model error). These updates ensure $\mathbb{D}(k) \subseteq \bar{\mathbb{D}}$ (compact set) for all $k \geq 0$, a critical requirement for proving uniform boundedness of the RPI set (Theorem \ref{theorem1}).
\begin{assumption}\label{assumption2}
The sequence of adapted disturbance sets $\{\mathbb{D}(k)\}_{k=0}^\infty$ satisfies: Each $\mathbb{D}(k)$ is compact and contains the origin. There exists a compact set $\bar{\mathbb{D}} \subset \mathbb{R}^n$ such that $\mathbb{D}(k) \subseteq \bar{\mathbb{D}}$ for all $k \geq 0$. The true disturbance satisfies $\bm{\Delta}^d_{vw}(k) \in \mathbb{D}(k)$ for all $k \geq 0$.
\end{assumption}
\begin{assumption}\label{assumption3}
 There exist a terminal cost function $V_f: \mathbb{R}^n \to \mathbb{R}_{\geq 0}$ and a terminal set $\mathbb{X}_f \subseteq \mathbb{X}_\mathbb{S}$ such that for all $\bm{x} \in \mathbb{X}_f$, there exists $\bm{u} = \kappa_f(\bm{x}) \in \mathbb{U}_\mathbb{S}$ satisfying: (i) $\hat{f}^d(\bm{x}, \kappa_f(\bm{x})) \in \mathbb{X}_f$, (ii) $V_f(\hat{f}^d(\bm{x}, \kappa_f(\bm{x}))) - V_f(\bm{x}) \leq -\ell(\bm{x}, \kappa_f(\bm{x}))$. 
    \begin{assumption}\label{assumption4}
        It is assumed that Problem \ref{5prob1} is feasible at time $k=0$.
    \end{assumption}
\end{assumption}
\begin{assumption}\label{assumption5}
    The discrete PIML model Jacobian $\tilde{\bm{A}}(k) = \frac{\partial \hat{f}^d}{\partial \hat{\bm{x}}} \big|_{(\hat{\bm{x}}(k), \hat{\bm{u}}(k))}$ is globally Lipschitz continuous on the compact state/input sets $\mathbb{X}\times \mathbb{U}$ with lispchitz constant $\mathcal{L}_A>0$: $$||\tilde{A}(k_1)-\tilde{A}(k_2)||\leq\mathcal{L}_A.(||\hat{x}(k_1)-\hat{x}(k_2)||+||\hat{u}(k_1)-\hat{u}(k_2)||),$$ for all $k_1, k_2 \geq 0$, which holds for quadrotor dynamics because rigid-body mechanics are smooth and the sparse PIML residual model adds only bounded nonlinearity-Lipschitz continuity is a direct consequence of smoothness on compact sets.
\end{assumption}
\begin{assumption}\label{assumption6}
    The PIML model coefficient update $\Delta\bm \xi(k)=\bm \xi(k+1)-\bm \xi(k)$ is uniformly bounded for all $k\geq0$:
    $$\Delta\bm \xi(k)\leq\bar{\Delta\bm \xi}, ~ \bar{\Delta\bm \xi}>0.$$ 
    Implemented in Algorithm 1 via a clipping step: if $||\Delta\bm \xi||>\bar{\Delta\bm \xi}$, set $\Delta\bm \xi(k)=\bar{\Delta\bm \xi}.\Delta\bm \xi(k)/||\Delta\bm \xi(k)||$. This assumption is physically meaningful for aerial vehicles-PIML model coefficients (aerodynamic residuals) evolve slowly with flight data, so large instantaneous updates are unphysical and indicative of noise.
\end{assumption}
\begin{assumption}\label{assumption7}
    The quadratic stage cost is Lipschitz continuous on $\mathbb{X}\times\mathbb{U}$ with lipschitz constant $\mathcal{L}_x>0$ and $\mathcal{L}_u>0$:
    \begin{equation}\begin{aligned}
    &|\ell_c(x_1,u_1)-\ell_c(x_2,u_2)|\leq\mathcal{L}_x.||x_1-x_2||+\mathcal{L}_u.||u_1-u_2||,\\& \forall(x_1,u_1),(x_2,u_2)\in \mathbb{X}\times\mathbb{U}. 
    \end{aligned}
    \end{equation}
    This trivially holds for quadratic costs on compact sets, proven by expanding the quadratic form and applying the Cauchy-Schwarz inequality (\ref{app:lipschitz_cost}).
\end{assumption}
\begin{assumption}\label{assumption8}
    The tube feedback gain $K\in \mathbb{R}^{4\times12}$ is uniformly bounded: $$||K||\leq\bar{K}, ~\bar{K}>0.$$ 
\end{assumption}
\begin{assumption}\label{assumption9}
    The quadrotor pitch angle $\theta$ (Euler angle) is constrained to a compact, non-singular operating region for all 
$k\geq0$: $\theta\in [-\theta_{max}, \theta_{max}], ~0<\theta_{max}<\pi/2$. Aerial vehicles (especially nano-quadrotors like the Crazyflie) do not operate in the Euler angle singular region ($\theta=\pm\pi/2$) for normal flight tasks (hovering, trajectory tracking). We set $\theta_{max}=\pi/4~(45^\circ)$ in all experiments, well within the non-singular region for quadrotor maneuvering.
\end{assumption}
With the PIML dynamics model and other preliminaries established, we now incorporate it into a robust MPC framework that can handle the residual uncertainties inherent in learned models.
\subsection{MPC Optimization problem}
At each time step $k$, given the current state $\bm x_k$, the following finite-horizon optimal control problem is solved:
\begin{problem}\label{5prob1}
\begin{subequations}
\label{5eq:optim}
\begin{align}
V_{N}(\bm x_k)= \underset{\hat{\bm{u}}\left(k\right)}{\mathrm{min}}
& \quad \sum_{j=0}^{N-1}\ell_c\left(\hat{\bm{x}}(j|k), \hat{\bm{u}}(j|k)\right)+V_{f}\left(\hat{\bm{x}}(N|k)\right), \label{5eq:cost}\\
\mathrm{subject~to} 
& \quad \hat{\bm{x}}\left(0 | k\right)=\bm{x}(k), \label{5eq:const1}\\
& \hat{\bm{x}}\left(j+1 | k\right)=\hat f^d\left(\hat{\bm{x}}\left(j | k\right), \hat{\bm{u}}\left(j | k\right)\right), \label{5eq:const2}\\
& \quad \hat{\bm{u}}\left(j | k\right) \in \mathbb{U}_\mathbb{S},\label{5eq:const2a}\\
& \quad \hat{\bm{x}}\left(j | k\right) \in \mathbb{X}_\mathbb{S},\label{5eq:const2ace}\\
& \quad \hat{\bm{x}}\left(N| k\right) \in \mathbb{X}_f\subseteq \mathbb{X}_\mathbb{S},\label{5eq:const2ace1}
\end{align}
\end{subequations}
where \eqref{5eq:const1}, \eqref{5eq:const2}, \eqref{5eq:const2a}, \eqref{5eq:const2ace}, \eqref{5eq:const2ace1} are the initial condition, dynamic constraints, input constraints, state constraints, and terminal constraints, respectively. Let $\hat{\bm{u}}^*(k) = {\hat{u}^*(0|k), \ldots, \hat{u}^*(N-1|k)}$ be the optimal solution. The applied control is: $\bm x_k=\hat{u}^* (0|k)+K(\bm x(k)-\hat{\bm x}^*(0|k))$.
\end{problem}
The Proposed algorithm is described in Algorithm \ref{algo:adaptive_mpc_piml}, and the schematic diagram of the presented scheme is depicted in Fig.~\ref{fig:system_architecture}. 
\begin{algorithm*}[htbp!]
\caption{Proposed algorithm}
\label{algo:adaptive_mpc_piml}
\SetAlgoLined
\KwIn{$\bm{x}(0)$, $\mathbb{D}(0)$, $T_{\text{learn}}$, $\lambda$, $\gamma$, $\bar{\Delta\bm \xi}$, $\bar{\Delta_{max}}$} 
\KwOut{Control inputs $\bm{u}(k)$ for $k = 0, 1, 2, \dots$}

Initialize PIML model $\dot{\hat{\bm{x}}} = \bm{\Psi}(\hat{\bm{x}}, \hat{\bm{u}})\bm{\xi}$ with $\bm{\xi} = \bm{0}$\;
Compute initial RPI set $\mathbb{S}(0)$ for $\tilde{\bm{A}}(0)$ and $\mathbb{D}(0)$\;
Set tightened constraints: $\mathbb{X}_\mathbb{S} = \mathbb{X} \ominus \mathbb{S}(0)$, $\mathbb{U}_\mathbb{S} = \mathbb{U} \ominus K\mathbb{S}(0)$\;
Initialize dataset $\mathcal{D} = \{\hat{\bm{x}}, \hat{\bm{u}}, \dot{\hat{\bm{x}}}\}$\;

\For{$k = 0, 1, 2, \dots$}{

    Solve Problem~\ref{5prob1}:
    Obtain $\hat{\bm{u}}^*(k)$ and state $\hat{\bm{x}}^*(k)$\;

    Compute tracking error: $\tilde{\bm{e}}(k) = \bm{x}(k)-\hat{\bm{x}}^*(0|k)$\;
    Apply control: $\bm{u}(k) = \hat{\bm{u}}^*(0|k) + K\tilde{\bm{e}}(k)$\;
 
    Measure next state $\bm{x}(k+1)$\;
    Compute realized disturbance in \eqref{step2_adaption}\;
    
    Add new data to dataset: $\mathcal{D} \leftarrow \mathcal{D} \cup \{\bm{x}(k), \bm{u}(k), \bm{\Delta}^d_{vw}(k)\}$\;
    
    \If{$k \mod T_{\text{learn}} = 0$ \textbf{and} $|\mathcal{D}| > N_{\min}$}{
    
        Construct library matrix $\bm{\Psi}(\hat{\bm{x}}, \hat{\bm{u}})$ from $\mathcal{D}$\;
        Solve sparse regression \eqref{5sindy5} for each state dimension $j$\;
        Update PIML model coefficients: $\Delta\bm{\xi} \leftarrow [\bm{S}_1, \bm{S}_2, \dots, \bm{S}_{12}]-\bm \xi$\;
        Clip parameter update (Assumption \ref{assumption6}): if $||\Delta\bm \xi||>\bar{\Delta\bm \xi}$, set $\Delta\bm \xi=\bar{\Delta\bm \xi}.\frac{\Delta\bm \xi}{||\Delta\bm \xi||}$\;
        Update PIML model coefficients: $\bm \xi\leftarrow\bm \xi+\bar{\Delta\bm \xi}$\;
        Update nominal dynamics: $\hat{f}^d \leftarrow \bm{\Psi}(\hat{\bm{x}}, \hat{\bm{u}})\bm{\xi}$\;
    }
    Update disturbance set center $c(k+1)$\;
    Update disturbance bounds:
    $\bar{\bm{\Delta}}^d_{vw}(k+1) = \gamma\bar{\bm{\Delta}}^d_{vw}(k)+(1-\gamma) ||\bm{\Delta}^d_{vw}(k) - c(k+1)||_{\infty}$\;
    Construct an adapted disturbance set $\mathbb{D}(k+1)$\;
    Compute Jacobian: $\tilde{\bm{A}}(k+1) = \frac{\partial \hat{f}^d}{\partial \hat{\bm{x}}} \big|_{(\hat{\bm{x}}(k), \hat{\bm{u}}(k))}$\;
    Compute learning-induced uncertainty: $\mathcal{U}_{\xi}(k+1)=\mathcal{L}.||\Delta\bm \xi||.\mathbb{B}_r$\;
    Compute new RPI set $\mathbb{S}(k+1)$ satisfying \eqref{eq14R}\;
    Update tightened constraints:
    $\mathbb{X}_\mathbb{S}(k+1) = \mathbb{X} \ominus \mathbb{S}(k+1), \quad \mathbb{U}_\mathbb{S}(k+1)= \mathbb{U} \ominus K\mathbb{S}(k+1)$\;
    Shift prediction horizon for next iteration\;
}
\end{algorithm*}
\begin{figure}[htbp!]
\centering
\begin{tikzpicture}[
    node distance=0.5cm and 0.5cm,
    >=stealth,
    block/.style={draw, fill=blue!10, rounded corners, minimum width=2cm, minimum height=1cm, align=center},
    initblock/.style={draw, fill=green!10, rounded corners, minimum width=2cm, minimum height=1cm, align=center},
    refblock/.style={draw, fill=purple!10, rounded corners, minimum width=2cm, minimum height=1cm, align=center},
    uavblock/.style={draw, fill=orange!10, rounded corners, minimum width=2cm, minimum height=1cm, align=center},
    sensorblock/.style={draw, fill=red!10, rounded corners, minimum width=2cm, minimum height=1cm, align=center},
    dataflow/.style={font=\small, align=center}
]

\node[initblock] (init) {\textbf{Initialization}};
\node[above=0.0001cm of init] {\scriptsize{Line 1- Line 4}};

\node[block, right=1.0cm of init] (algorithm) {\textbf{PIML-MPC}};
\node[above=0.0001cm of algorithm] {\scriptsize{Line 5- Line 28}};

\node[uavblock, right=1.5cm of algorithm] (uav) {\textbf{UAV}};

\node[sensorblock, below=0.8cm of uav] (sensors) {\textbf{Feedback}};

\draw[->, thick, black] (init) -- node[above] {} (algorithm);
\draw[->, thick, black] (algorithm) -- node[above, dataflow] {$\bm{u}(k)$} (uav);
\draw[->, thick, black] (uav) -- node[right, dataflow, pos=0.5] {$\bm{x}(k)$} (sensors);
\draw[->, thick, black] (sensors.west) -| node[below, dataflow, pos=0.25] {$\bm{x}(k)$} ([yshift=0cm]algorithm.south);

\node[above=0.8cm of uav] (distlabel) {};
\draw[->, thick, black] (distlabel) -- node[right, dataflow, pos=0.6] {$\bm\Delta_{vw}$} (uav);
\end{tikzpicture}
\caption{Overall architecture of the control strategy (read with Algorithm \ref{algo:adaptive_mpc_piml})}
\label{fig:system_architecture}
\end{figure}
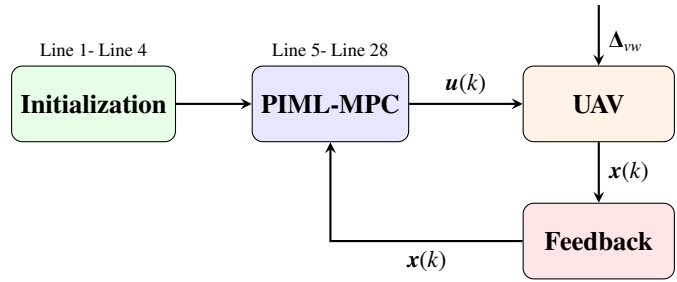
\begin{remark}
The two computationally intensive steps in Algorithm \ref{algo:adaptive_mpc_piml} are: (1) Sparse regression for automatic library pruning and coefficient learning, and (2) Recursive computation of the time-varying RPI set. Sparse regression operates on a compact, automatically generated library (not an excessively large dictionary) and is executed only periodically (every $T_{\text{learn}}$ steps, not at every time step), which keeps its computational cost low and compatible with embedded hardware. RPI set computation uses a recursive, linear-matrix-based update rule rather than complex convex optimization or polytope propagation, resulting in constant and predictable computational complexity per update. Thus, despite involving learning and robust set operations, the overall computational load remains manageable for real-time quadrotor control, as validated experimentally in Section \ref{section4}.
\end{remark}
We now provide theoretical guarantees for this proposed framework.
\section{Theoretical guarantees}\label{section3}
\subsection{Feasibility}
\begin{theorem}\label{theorem1}
    Under Assumptions \ref{assumption2}-\ref{assumption6}, if the Problem \ref{5prob1} is feasible at time $k$, then it is feasible at all future times. Moreover, the time-varying RPI set $\mathbb{S}(k)$ and tightened constraint sets $\mathbb{X}_\mathbb{S}(k)$, $\mathbb{U}_\mathbb{S}(k)$ are uniformly bounded for all $k\geq0$.
    \begin{proof}
   We proceed by induction and construction of a feasible candidate solution for Problem \ref{5prob1} at time $k+1$, then use uniform boundedness of the RPI set to verify constraint satisfaction. In the Step 1 related to induction hypothesis, assume Problem \ref{5prob1} is feasible at time $k$, with optimal nominal state/input sequences $\hat{\bm{x}}^*(j|k) = \mathbb{X}_{\mathbb{S}}(k)$ and $\hat{\bm{u}}^*(j|k) = \mathbb{U}_{\mathbb{S}}(k)$ for $j=0,1,\ldots, N-1$, and terminal state $\hat{\bm{x}}^*(N|k) \in \mathbb{X}_{f}\subseteq \mathbb{X}_{\mathbb{S}}(k)$. The applied control at time $k$ is $\bm u(k) = \hat{\bm u}^*(0|k) + K \tilde{\bm e}(k)$, with $\tilde{\bm e}(k) \in \mathbb{S}(k)$. In the Step 2 for candidate solution at $k+1$, construct a candidate solution $(\tilde{\bm x}(j|k+1), \tilde{\bm u}(j|k+1))$ for Problem 1 at time $k+1$:
\begin{enumerate}
    \item Shifted control sequence: $\tilde{\bm u}(j|k+1) = \hat{\bm u}^*(j+1|k)$ for $j=0,\dots,N-2$.
    \item Terminal control: $\tilde{\bm u}(N-1|k+1) = \kappa_f(\tilde{\bm x}(N-1|k+1))$ (Assumption \ref{assumption3}).
    \item Candidate state sequence: $\tilde{\bm x}(0|k+1) = \bm x(k+1)$ (initial condition), $\tilde{\bm x}(j+1|k+1) = \hat{f}^d(\tilde{\bm x}(j|k+1), \tilde{\bm u}(j|k+1))$ for $j=0,\dots,N-1$, and $\tilde{\bm x}(N|k+1) = \hat{f}^d(\tilde{\bm x}(N-1|k+1), \kappa_f(\tilde{\bm x}(N-1|k+1)))$.
\end{enumerate}
 In Step 3 for uniform boundedness of $\mathbb{S}(k)$: From the time-varying RPI set condition \eqref{eq14R} and Assumptions \ref{assumption2}, \ref{assumption5}, \ref{assumption6}:
 \begin{itemize}
    \item $\mathbb{D}(k) \subseteq \bar{\mathbb{D}}$ (compact, Assumption \ref{assumption2}),
    \item $\mathcal{U}_{\xi}(k) = \mathcal{L}_{\xi} ||\Delta\bm \xi(k)|| \mathbb{B}_r \subseteq \mathcal{L}_\xi \bar{\Delta\bm \xi} \mathbb{B}_r = \bar{\mathcal{U}_\xi}$ (compact, Assumption \ref{assumption6}),
    \item $\tilde{A}(k)$ is Lipschitz continuous (Assumption \ref{assumption5}) and thus bounded on $\mathbb{X} \times \mathbb{U}$ (compact set).
\end{itemize}
   By the Bounded RPI Set Lemma (standard result \citep{Limon2009}), the time-varying RPI set $\mathbb{S}(k)$ is uniformly bounded for all $k\geq 0$: there exists a compact set $\bar{\mathbb{S}}\subset\mathbb{R}^{12}$ such that $\mathbb{S}(k)\subseteq \bar{\mathbb{S}}$ for all $k\geq0$. 

In Step 4, the tightened constraint sets are defined as $\mathbb{X}_{\mathbb{S}}(k) = \mathbb{X} \ominus \mathbb{S}(k)$ and $\mathbb{U}_{\mathbb{S}}(k) = \mathbb{U} \ominus K \mathbb{S}{k}$. Since $\mathbb{S}{k} \subseteq \bar{\mathbb{S}}$ (uniformly bounded), the tightened sets satisfy:
\[
\mathbb{X}_{\mathbb{S}}(k) \supseteq \mathbb{X} \ominus \bar{\mathbb{S}} = \bar{\mathbb{X}}_{\mathbb{S}}, \quad \mathbb{U}_{\mathbb{S}}(k) \supseteq \mathbb{U} \ominus K \bar{\mathbb{S}} = \bar{\mathbb{U}}_{\mathbb{S}},
\]
where $\bar{\mathbb{X}}_{\mathbb{S}}, \bar{\mathbb{U}}_{\mathbb{S}}$ are non-empty compact sets (Pontryagin difference of compact sets is compact/non-empty if $\mathbb{S}(k)$ contains the origin-Assumption \ref{assumption2}). In Step 5, the candidate solution satisfies all constraints.
\begin{itemize}
    \item Initial Condition \eqref{5eq:const1}: $\tilde{\bm x}(0|k+1) = \bm x(k+1)$ (by construction)-satisfied.
    \item Dynamic Constraint \eqref{5eq:const2}: $\tilde{\bm x}(j+1|k+1) = \hat{f}^d(\tilde{\bm x}(j|k+1), \tilde{\bm u}(j|k+1))$ (by construction)-satisfied.
    \item Input/State Constraints \eqref{5eq:const2a}, \eqref{5eq:const2ace}: For $j=0,\dots,N-2$, $\tilde{\bm u}(j|k+1) = \hat{\bm u}^*(j+1|k) \in \mathbb{U}_{\mathbb{S}}(k) \subseteq \bar{\mathbb{U}}_{\mathbb{S}} \subseteq \mathbb{U}_{\mathbb{S}}(k+1)$ and $\tilde{\bm x}(j|k+1) = \hat{\bm x}^*(j+1|k) \in \mathbb{X}_{\mathbb{S}}(k) \subseteq \bar{\mathbb{X}}_{\mathbb{S}} \subseteq \mathbb{X}_{\mathbb{S}}(k+1)$ (from uniform boundedness of $\mathbb{S}(k)$)-satisfied.
    \item Terminal Constraint \eqref{5eq:const2ace1}: $\tilde{\bm x}(N-1|k+1) = \hat{\bm x}^*(N|k) \in \mathbb{X}_f$ (induction hypothesis). By Assumption \ref{assumption3}, $\tilde{\bm x}(N|k+1) = \hat{f}^d(\tilde{\bm x}(N-1|k+1), \kappa_f(\tilde{\bm x}(N-1|k+1))) \in \mathbb{X}_f \subseteq \mathbb{X}_{\mathbb{S}}(k+1)$-satisfied.
\end{itemize}
The candidate solution satisfies all constraints of Problem \ref{5prob1} at time $k+1$, so Problem 1 is feasible at $k+1$. By Assumption \ref{assumption4} (feasibility at $k=0$), Problem \ref{5prob1} is feasible for all $k \geq 0$.   
\end{proof}
\end{theorem}
\subsection{Stability}
\begin{theorem}\label{theorem2}
    If the Assumptions \ref{assumption2}-\ref{assumption8} hold, the closed-loop system consisting of the PIML model, error dynamics, and MPC control law is ISS w.r.t. the augmented disturbance input $[\bm{\Delta}^d_{vw}(k)^\top~\Delta_{\xi}(k)^\top]^\top \in \mathbb{R}^{24}$. Specifically, there exist $\mathcal{K}_{\infty}$-functions $\alpha_1$, $\alpha_2$, $\alpha_3$ and a $\mathcal{K}$-function $\sigma$ such that for all initial states $\bm{x}(0)\in \mathbb{X}$ and all bounded augmented disturbance sequences ${\bm \Delta_{aug}(k)}^{\infty}_{k=0}$, the closed-loop state $\bm{x}(k)$ satisfies:
    \begin{equation}\label{20stability}
        \alpha_{1}(||\bm x(k)||)\leq V^{*}_{N}(\bm x(k))\leq\alpha_2(||\bm x(k)||),
    \end{equation}
    \begin{equation}\label{21stability}
       V^{*}_{N}(\bm x(k+1))-V^{*}_{N}(\bm x(k))\leq\alpha_3(||\bm x(k)||)+\sigma(||\bm \Delta_{aug}(k)||),
    \end{equation}
    where $V^{*}_{N}(\bm x(k))$ is the optimal value function of Problem \ref{5prob1} at time $k$.
    \begin{proof}
    We proceed in four key steps: (1) verify Lyapunov function properties \eqref{20stability}, (2) derive nominal Lyapunov descent, (3) extend to the disturbed/learned system \eqref{21stability}, and (4) conclude ISS stability. In Step 1, we discuss the Lyapunov Function Properties \eqref{20stability}. The optimal value function is defined as:
\[
V_N^*(\bm x(k)) = \min_{\hat{\bm u}(\cdot)} \sum_{j=0}^{N-1} \ell_c(\hat{\bm x}(j|k),\hat{\bm u}(j|k)) + V_f(\hat{\bm x}(N|k)),
\]
where $\ell_c(\bm x,\bm u) = (\bm x-\bm x_{\text{ref}})^\top \bm Q (\bm x-\bm x_{\text{ref}}) + (\bm u-\bm u_{\text{ref}})^\top \bm R (\bm u-\bm u_{\text{ref}})$ (quadratic stage cost, $\bm Q \succeq 0$, $\bm R \succ 0$) and $V_f(x)$ is the terminal cost (Assumption~\ref{assumption3}).
\begin{itemize}
    \item Lower Bound ($\alpha_1$): 
      Follows from non-negativity of $\ell_c(\bm x,\bm u)$ and $V_f(x)$. Define $\alpha_1(r) = c_1 r^2$ (a $\mathcal{K}_\infty$-function) with $c_1 = \lambda_{\text{min}}(\bm Q)/2$ (minimum eigenvalue of $\bm Q$). This gives:
      \[
      V_N^*(\bm x(k)) \geq \alpha_1(||\bm x(k)||).
      \]
    \item Upper Bound ($\alpha_2$): Follows from compactness of $\mathbb{X} \times \mathbb{U}$ and Lipschitz continuity of $\ell_c(\bm x,\bm u)$ (Assumption \ref{assumption7}). Define $\alpha_2(r) = c_2 r^2 + c_3$ (a $\mathcal{K}_\infty$-function) with $c_2 = \lambda_{\text{max}}(\bm Q) + \lambda_{\text{max}}(\bm R)$ (maximum eigenvalues) and $c_3$ a constant from the terminal cost $V_f(x)$. This gives: $V_N^*(\bm x(k)) \leq \alpha_2(\bm x(k))$.
\end{itemize}
Both $\alpha_1, \alpha_2$ are valid $\mathcal{K}_\infty$-functions (strictly increasing, continuous, $\alpha_i(0)=0$, $\alpha_i(r) \to \infty$ as $r \to \infty$), so \eqref{20stability} holds. In Step 2, for nominal Lyapunov descent, for the nominal system (no disturbance/learning uncertainty: $\bm \Delta_{aug}(k) = 0$), the optimal value function satisfies the standard MPC descent condition:
\begin{equation}\label{22stab}
   V_N^*(\bm x(k+1)) - V_N^*(\bm x(k)) \leq -\ell_c(\hat{\bm x}^*(0|k), \hat{\bm u}^*(0|k)). 
\end{equation}
By the quadratic stage cost, $\ell_c(\hat{\bm x}^*(0|k), \hat{\bm u}^*(0|k)) \geq \alpha_3(\bm x(k))$ where $\alpha_3(r) = c_1 r^2$ (a $\mathcal{K}_\infty$-function)-this is the nominal stability condition. In Step 3, for the extension to the disturbed/learned System \eqref{21stability}, we extend \eqref{22stab} to the actual closed-loop system (applied control: $\bm u(k) = \hat{\bm u}^*(0|k) + K \tilde{\bm e}(k)$, $\tilde{\bm e}(k) = \bm x(k)-\hat{\bm x}(k)$):
\begin{enumerate}
    \item State Error Bound:  The tracking error $\tilde{\bm e}(k) \in \mathbb{S}(k)$ (RPI set), and $\mathbb{S}(k) \subseteq \bar{\mathbb{S}}$ (uniformly bounded, Theorem \ref{theorem1}). Thus:
      \begin{equation}\label{23stab}
           ||\bm x(k) - \hat{\bm x}^*(0|k)|| = ||\tilde{\bm e}(k)|| \leq ||\bar{\mathbb{S}}|| = C_S, 
      \end{equation}
      where $C_S > 0$ is a constant (RPI set bound).
    \item Control Error Bound: From Assumption \ref{assumption8} ($||K|| \leq \bar{K}$, bounded LQR feedback gain):
      \begin{equation}\label{24stab}
      ||\bm u(k) - \hat{\bm u}^*(0|k)|| = ||K \tilde{\bm e}(k)|| \leq \bar{K} C_S = C_K, 
      \end{equation}
      where $C_K > 0$ is a constant (control error bound).
    \item Lipschitz Cost Bound: By Assumption \ref{assumption7} (Lipschitz continuity of $\ell_c(\bm x,\bm u)$) and \eqref{23stab}-\eqref{24stab}:
      \begin{equation}\label{25stab}
      |\ell_c(\bm x(k),\bm u(k)) - \ell_c(\hat{\bm x}^*(0|k), \hat{\bm u}^*(0|k))| \leq \mathcal{L}_x C_S + \mathcal{L}_u C_K = C_\ell, 
       \end{equation}
      where $\mathcal{L}_x, \mathcal{L}_u$ are Lipschitz constants (Assumption 7) and $C_\ell > 0$ is a constant. 
    \item Augmented Disturbance Bound: The augmented disturbance $\bm \Delta_{aug}(k)$ is bounded (Assumptions \ref{assumption2}, \ref{assumption6}: $||\bm \Delta_{aug}(k)|| \leq C_{aug}$). Define the $\mathcal{K}$-function $\sigma(r) = C_\ell \cdot r$ (continuous, strictly increasing, $\sigma(0)=0$) to bound disturbance effects.
     \item ISS Descent Condition: Combining \eqref{22stab}-\eqref{25stab}:
      \begin{equation}
          \begin{aligned}
            &  V_N^*(\bm x(k+1)) - V_N^*(\bm x(k)) \leq -\ell_c(\bm x(k),\bm u(k)) + C_\ell\\& \leq -\alpha_3(||x(k)||) + \sigma(||\Delta_{aug}(k)||).
          \end{aligned}
      \end{equation}
\end{enumerate}
This is exactly \eqref{21stability}-the ISS descent condition for the augmented disturbance input. In Step 4 for ISS Conclusion, and by the ISS Lyapunov Theorem \citep{Sontag2008}, \eqref{20stability}-\eqref{21stability} are sufficient to conclude that the closed-loop system is ISS with respect to $\bm \Delta_{aug}(k)$. Boundedness of $\bm \Delta_{aug}(k)$ implies boundedness of the closed-loop state $\bm x(k)$-a critical robustness property for aerial vehicle control.
\end{proof}
\end{theorem}
\subsection{Robustness to Periodic PIML Model Updates}
The PIML model is updated periodically (every $T_{\text{learn}}$ time steps) via sparse regression, a discrete update that introduces a small perturbation to the nominal dynamics $\hat{f}^d$. A key theoretical question is whether periodic PIML model updates break the recursive feasibility (Theorem \ref{theorem1}) or ISS stability (Theorem \ref{theorem2}) guarantees, which we address with two lemmas that prove robustness to periodic updates (feasibility/stability are preserved) and a practical tuning rule for $T_{\text{learn}}$.
\begin{lemma}
    Under Assumptions \ref{assumption2}-\ref{assumption6} and Theorem \ref{theorem1}, periodic PIML model updates (every 
$T_{\text{learn}}$ steps) do not break the recursive feasibility of Problem \ref{5prob1}.
\begin{proof}
    PIML model updates change the nominal dynamics $\hat{f}^d\rightarrow\hat{f}_{new}^d$ with a bounded model error:
    \begin{equation}\label{eq26feas}
        ||\hat{f}_{new}^d(x,u)-\hat{f}^d(x,u)||\leq\mathcal{L}_{\xi}.||\Delta\bm \xi||\leq\mathcal{L}_{\xi}\bar{\Delta\bm \xi}=C_f,
    \end{equation}
where $C_f>0$ is a constant (from Assumption \ref{assumption6} and the Lipschitz constant $\mathcal{L}_\xi$). This model error is absorbed into the learning-induced uncertainty set $\mathcal{U}_\xi(k)$, which is included in the time-varying RPI set condition \eqref{eq14R}. Since RPI set $\mathbb{S}(k)$ is s uniformly bounded (Theorem \ref{theorem1}), the tightened constraint sets $\mathbb{X}_{\mathbb{S}}(k)$, $\mathbb{U}_{\mathbb{S}}(k)$ remain non-empty. Hence, recursive feasibility is preserved.
\end{proof}
\end{lemma}
\begin{lemma}
    Under Assumptions \ref{assumption2}–\ref{assumption8} and Theorem \ref{theorem2}, periodic PIML model updates (every $T_{\text{learn}}$) steps) do not break ISS stability of the closed-loop system.
    \begin{proof}
        Periodic model updates introduce a small perturbation to the ISS Lyapunov function $V_N^*(\bm x(k))$. The perturbation magnitude is bounded by the model error $C_f$ \eqref{eq26feas} and Lipschitz continuity of the stage cost (Assumption \ref{assumption7}):
        \begin{equation}
            |V_N^*(\bm x(k+T_{\text{learn}}))-V_N^*(\bm x(k))|\leq T_{\text{learn}} C_\ell C_f =C_p.
        \end{equation}
        where $C_p$ is a constant (perturbation bound). This perturbation is sublinear in $k$ and thus does not violate the ISS descent condition \eqref{21stability}, the $\mathcal{K}$-function $\sigma$ is simply extended to include $C_p(\sigma(r_l))=C_\ell r_l + C_p$, which remains a valid $\mathcal{K}$-function. ISS stability is preserved.
    \end{proof}
\end{lemma}
\begin{remark}
    For aerial vehicles, $T_{\text{learn}}$ is tuned to be slower than the MPC control loop (100–200 Hz) to minimize transients from model updates. We recommend: $T_{\text{learn}}=5-10 ~\text{time-steps}$ with a real-time duration of 50-100 ms, which corresponds to a PIML update frequency of 10–20 Hz (well below the control loop frequency). This tuning balances fast model learning (to adapt to unmodeled aerodynamics) and slow updates (to preserve closed-loop robustness).
\end{remark}
With theoretical guarantees established, we now validate the proposed approach through both numerical simulations and hardware experiments.
\section{Simulation and hardware experiments}\label{section4}
\begin{figure*}[htbp!]
    \centering
    \includegraphics[width=13cm, height=2.5cm]{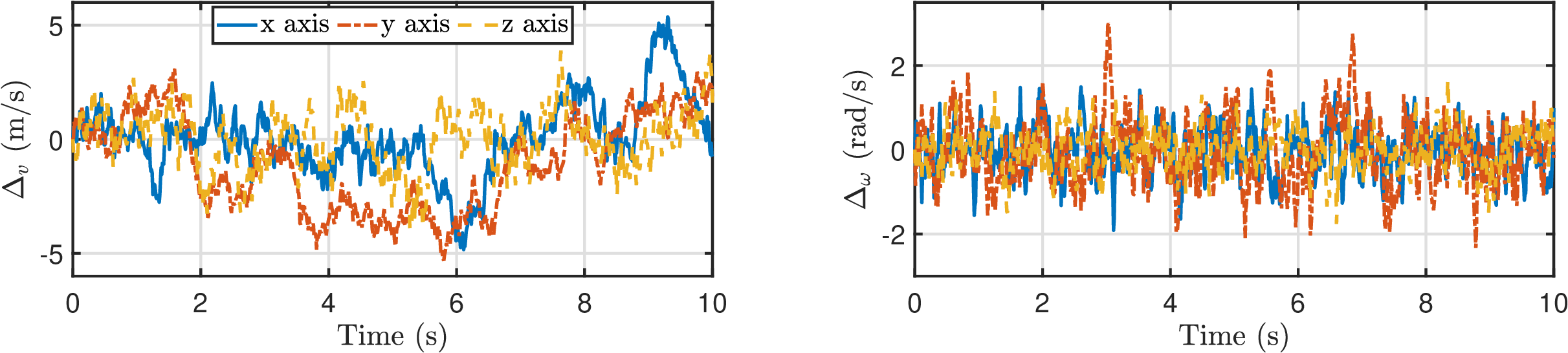}
    \caption{Disturbance profile used in numerical results}
    \label{fig:5}
\end{figure*}

The effectiveness of our PIML-MPC framework is demonstrated through comprehensive testing, beginning with simulation studies and proceeding to real-world implementation.
We executed the results of the proposed technique using MATLAB 2024b, transcribed by CasADi~\citep{Andersson2019}, version 3.7.2, and IPOPT~\citep{watcher}, version 3.14.17, on an AMD Ryzen 9 7940HS, 8-core processor with a speed of 3.1-5.2 GHz and 64 GB of RAM, running Ubuntu 25.10. The quadrotor parameters used are: mass $m=0.027$ kg, gravitational acceleration $g=9.81$ m/s$^2$, inertia matrix $\bm{J}=\text{diag}([1.4\times10^{-5}, 1.4\times10^{-5}, 2.17\times10^{-5}])$ kg$\cdot$m$^2$, arm length $l=0.046$ m, and torque-to-thrust ratio $\kappa=3.15\times10^{-3}$ Nm/N. The MPC horizon was set to $N=15$. The weighting matrices were $\bm{Q}=\text{diag}([10,10,10,5,5,5,2,2,2,1,1,1])$ and $\bm{R}=\text{diag}([0.1,0.1,0.1,0.1])$. The constraints are: $-0.5<\mathcal{P}_x, \mathcal{P}_y<0.5$, $1.5<\mathcal{P}_z<2.5$, $-1<v_x, v_y<1$, $-2.5<v_y<2.5$, $-1<\phi, \theta<1$, $-0.5<\psi<0.5$. The input constraints reflect actuator limitations: $0 < u_1 < 0.4 $, $-0.02 < u_2, u_3, u_4 < 0.02$. The disturbance $\bm{\Delta}_{vw}(t)$ is Dryden Wind Turbulence model MIL-HDBK-1797B, as shown in Fig.~\ref{fig:5}. This profile represents challenging conditions that test the robustness of each controller. 
Four benchmark controllers were implemented for comparison: 1) Standard cascaded PID architecture with inner-loop rate control and outer-loop position control. Both loops employed a high-fidelity PIM; 2) Nonlinear MPC (SMPC): Traditional SMPC using the first-principles model without learning or robustness mechanisms. Used the same horizon and weights as our proposed method for fair comparison; 3) Fixed-Tube MPC (FT-MPC): Robust MPC with constant disturbance bound $\mathbb{D}_{\text{max}}$ covering 95\% of expected disturbances. The tube size was fixed based on worst-case disturbance estimates; 4) NN-MPC: MPC with a 3-layer NN-based learned model from the same dataset, serving as a black-box learning baseline. Both PID and SMPC employed a high-fidelity PIM. To ensure fair and reproducible comparison, we provide precise definitions of all benchmark controllers (PID, SMPC, FT-MPC, NN-MPC) used in simulations and hardware experiments:\\ (1) \textbf{PID Controller}: A standard cascaded PID architecture is implemented:
\[
\bm u_{\text{PID}}(k) = K_p e(k) + K_i \sum_{t=0}^k \bm e(t) \Delta t + K_d \frac{\bm e(k) - \bm e(k-1)}{\Delta t},
\]
where $\bm e(k) = \bm x_{\text{ref}}(k) - \bm x(k)$ is the tracking error, $\Delta t = 0.01$ s (sampling time), and tuning parameters (optimized via Ziegler-Nichols + manual refinement) in the outer loop (position/attitude): $$\bm{K}_{p,i,d} = ([4, 4, 6, 150, 150, 80],[0.5, 0.5, 1, 20, 20, 10], [0.8, 0.8, 1.2, 2, 2, 1.5]),$$ and in the inner loop (velocity/angular rate): $$\bm{K}_{p,i,d} = ([20, 20, 15, 30, 30, 25],[5, 5, 3, 5, 5, 4], [2, 2, 1.5, 0.5, 0.5, 0.3]).$$ (2) \textbf{SMPC}: Nominal MPC without disturbance learning, formulated as:
\[
\min_{\hat{u}(0|k),...,\hat{u}(N-1|k)} \sum_{j=0}^{N-1} \ell_c(\hat{x}(j|k),\hat{u}(j|k)) + V_f(\hat{x}(N|k)),
\]
subject to: $\hat{x}(j+1|k) = \hat{f}^d(\hat{x}(j|k),\hat{u}(j|k))$, $\hat{x}(j|k) \in \mathbb{X} = \{x \in \mathbb{R}^{12} : |x_i| \leq x_{\text{max},i}\}$ (same as proposed method), $\hat{u}(j|k) \in \mathbb{U} = \{u \in \mathbb{R}^4 : |u_i| \leq u_{\text{max},i}\}$ (same as proposed method).
Key parameters: Prediction horizon $N=15$, stage cost $\ell_c(x,u) = \bm x^\top \bm Q \bm x + \bm u^\top \bm R \bm u$ ($\bm Q = \text{diag}([10,...,1])$, $\bm R = \text{diag}([0.1,...,0.1])$), terminal cost $V_f(x) = x^\top \bm P x$. \\ (3) \textbf{FT-MPC}: RMPC with static disturbance bounds (no adaptation): Disturbance set: $\mathbb{D}_{\text{fixed}} = \{ \Delta \in \mathbb{R}^{12}: \|\Delta\|_\infty \leq 0.1 \}$ (tuned to 95\% of Dryden wind model bounds), Tube constraints: $\tilde{e}(j|k) = x(j|k) - \hat{x}(j|k) \in \mathbb{S}_{\text{fixed}}$ (static RPI set computed via [28]). Other parameters (horizon, cost, constraints) match SMPC for fairness.\\ (4) \textbf{NN-MPC}: MPC with dynamics modeled by a 3-layer feedforward NN: Input (12D state + 4D input) → Hidden Layer 1 (64 neurons, ReLU) → Hidden Layer 2 (32 neurons, ReLU) → Output (12D next state), Adam optimizer ($lr=0.001$), batch size=256, 10,000 epochs, training dataset=10,000 samples (same as proposed PIML dataset), MPC formulation: Identical to SMPC (horizon $N=15$, cost weights $\bm Q,\bm R$) but uses NN-predicted dynamics $\hat{x}(j+1|k) = \text{NN}(\hat{x}(j|k),\hat{u}(j|k))$.

All controllers are tuned to minimize tracking error while ensuring stability: PID: Ziegler-Nichols method + manual refinement (5 iterations of flight tests), SMPC/FT-MPC: Cost weights $\bm Q, \bm R$ tuned via grid search (100 combinations) to balance tracking and input smoothness, NN-MPC: Hyperparameter tuning (learning rate, batch size) via validation loss (early stopping at validation $\text{loss} <10^{-4}$), Proposed Method: Fixed hyperparameters ($\lambda=0.9$, $\gamma=0.95$, $\epsilon=10^{-4}$) across all experiments (no dataset-specific tuning).

\begin{figure*}
     \includegraphics[width=.3\textwidth]{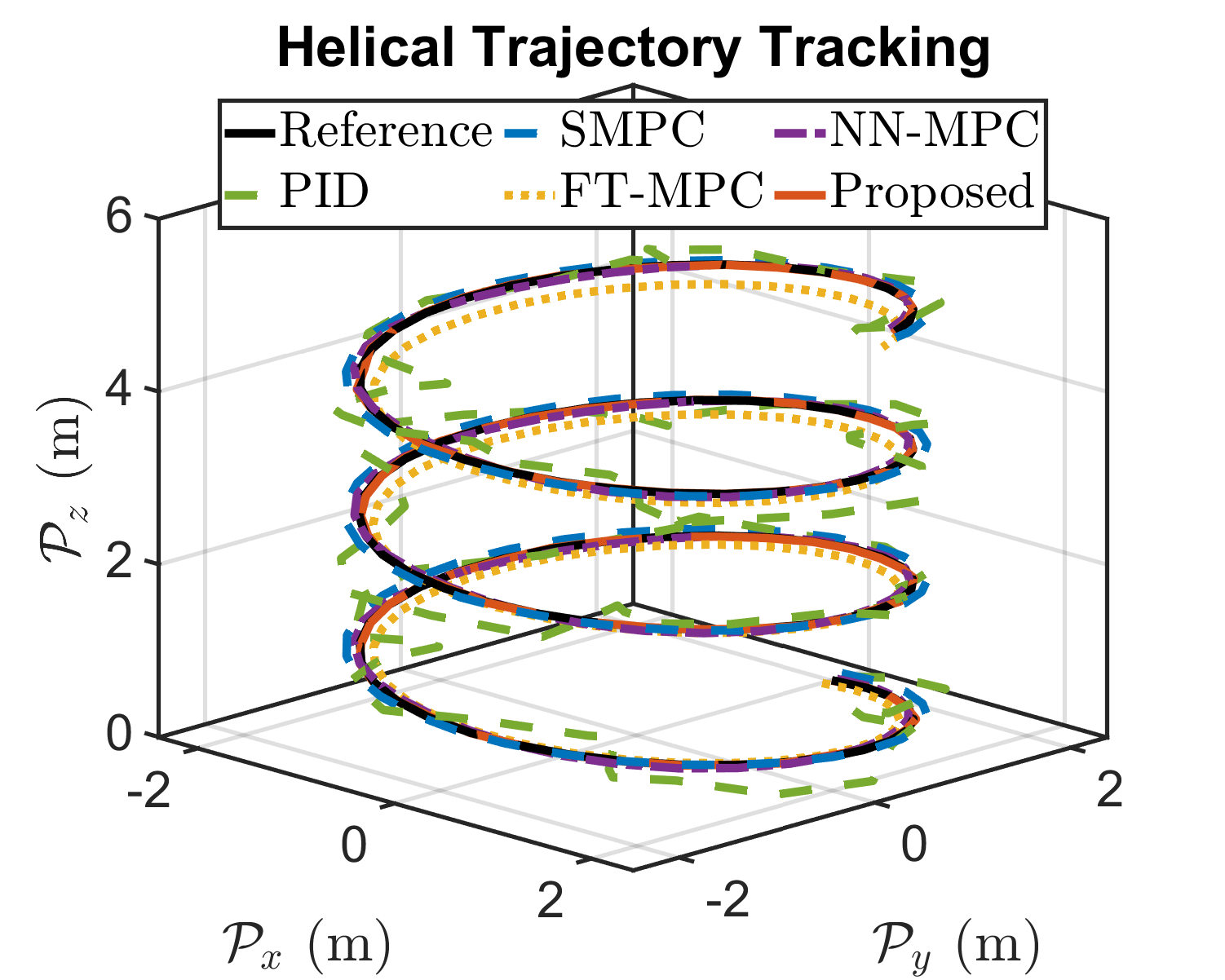}
    \hfill
    \includegraphics[width=.3\textwidth]{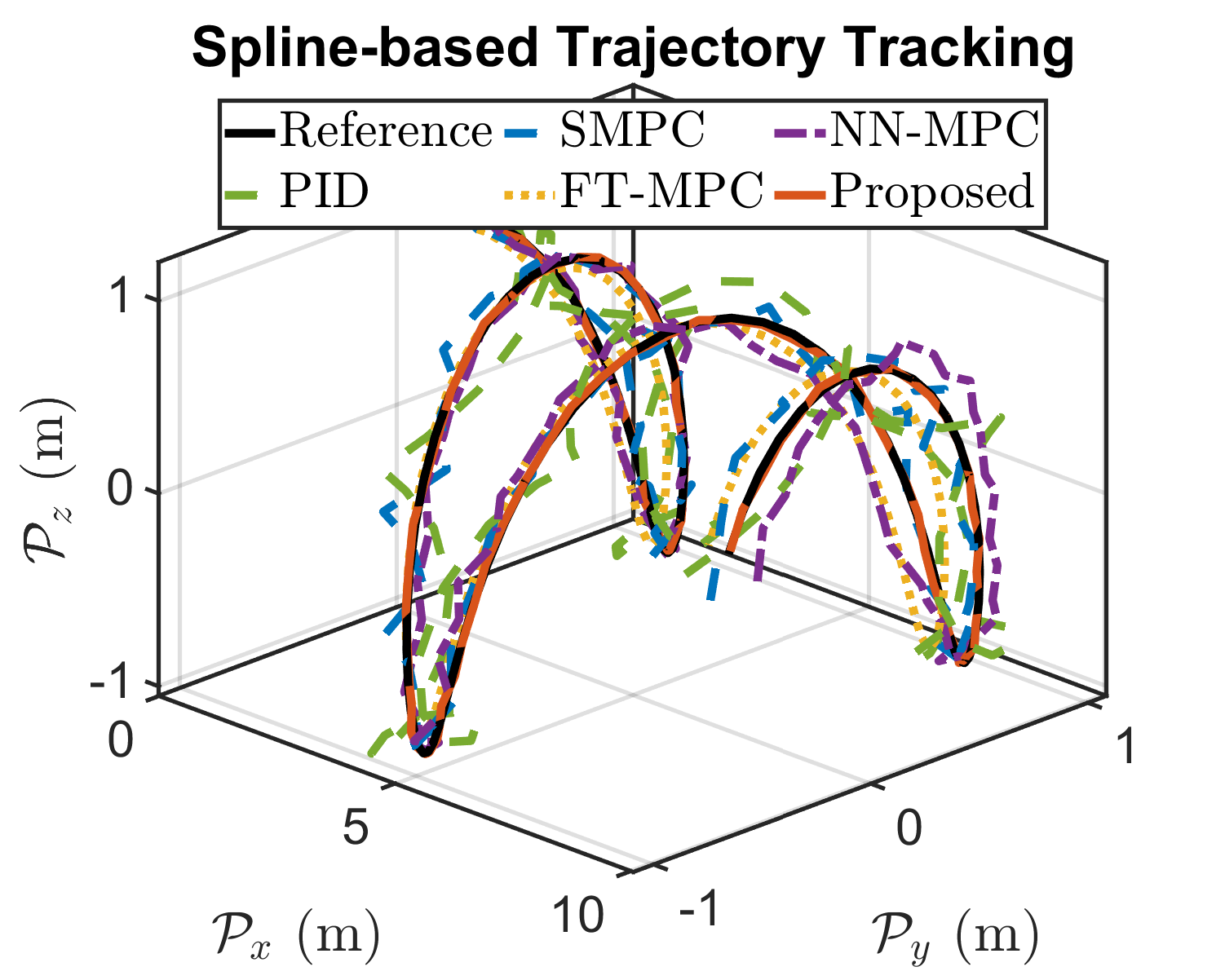}
    \hfill
    \includegraphics[width=.3\textwidth]{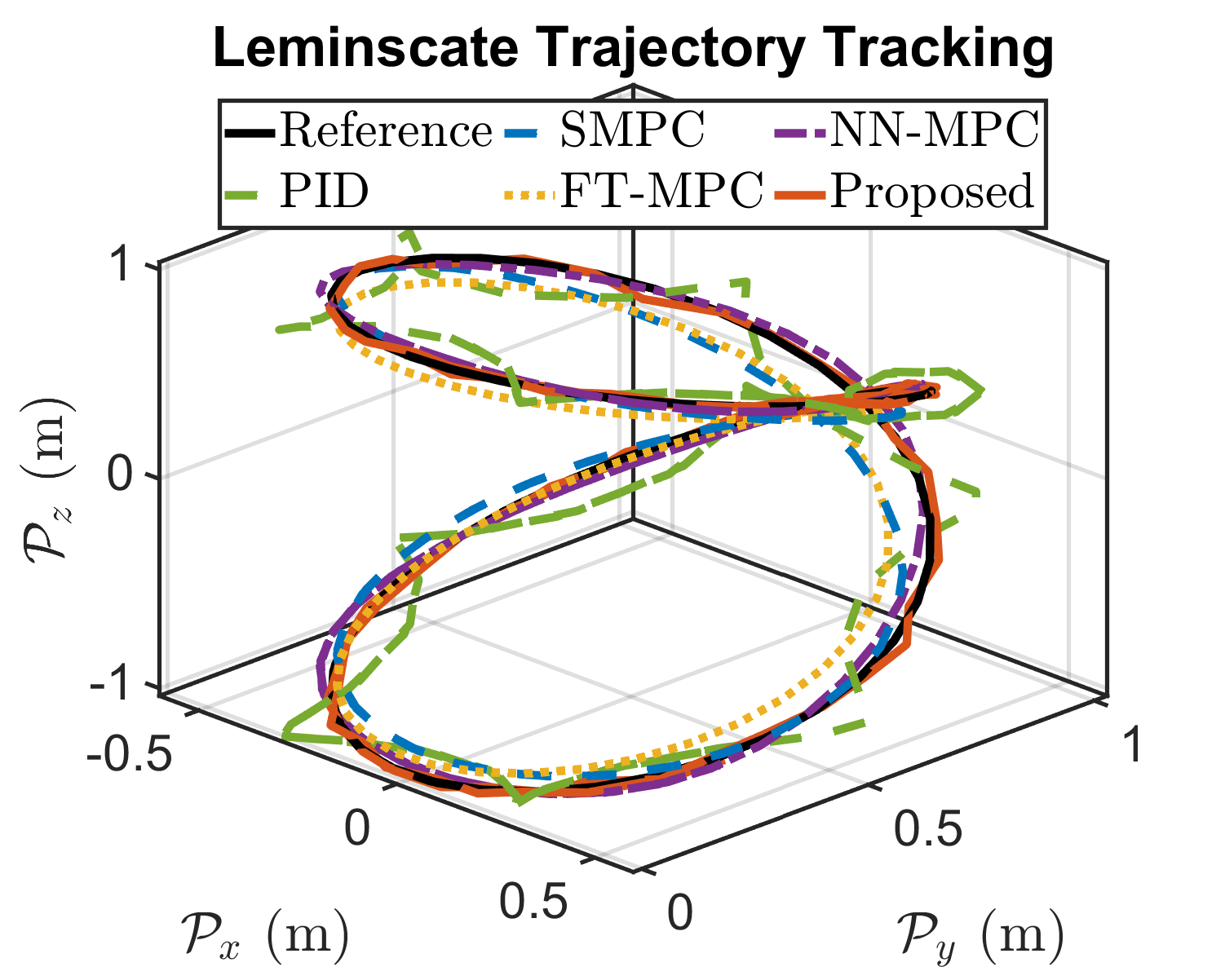}
    \caption{Trajectory tracking response for different types of trajectories}
    \label{fig:6}
\end{figure*}

Three diverse reference trajectories were employed to comprehensively evaluate controller performance under different maneuver characteristics: helical trajectory, spline-based trajectory, and lemniscate trajectory. The reference 3D trajectories $[\mathcal{P}_x~\mathcal{P}_y~\mathcal{P}_z]^\top$ for $t\in[0~10]$ are Helical trajectory $[t~~s_t+0.1s_{3t}~~c_{2t}]^\top$, Spline-based trajectory $[2s_t~~2c_{2t}~~0.5t]^\top$, Leminscate trajectory $[s_tc_t~~s_ts_t~~s_tc_t]^\top$. The tracking response is shown in Fig.~\ref{fig:6}. 
Also, Table~\ref{tab:tracking_errors} describes the performance analysis based on root mean square (RMSE). Our PIML-MPC achieves an optimal balance between computation time and tracking accuracy. Moreover, our method is 37.4\% faster than NN-MPC while achieving significantly effective tracking performance, demonstrating the efficiency of sparse PIML models over black-box NNs. To further quantitatively evaluate the tracking accuracy, stability, and robustness of all methods against random external uncertainties, we perform 20 independent Monte Carlo simulation trials for each control strategy. In each randomized trial, small bounded perturbations are imposed on the initial flight position, and stochastic time-varying wind disturbances are generated to simulate realistic uncertain flight scenarios. Based on the collected tracking RMSE data of all repeated experiments, we compute the average tracking error, standard deviation, and 95\% confidence interval (CI) for each controller. The comprehensive statistical indices are summarized in Table \ref{tab:stat_rmse}. Different from single-run deterministic results, these statistical outcomes quantitatively characterize the average tracking precision and error fluctuation degree of each controller in uncertain environments, significantly improving the rigor and persuasiveness of the comparative analysis.
    \begin{table}[htbp!]
    \centering
    \footnotesize
    \caption{Statistical RMSE over 20 randomized Monte Carlo trials (Unit: m)}
    \begin{tabular}{lcccc}
    \hline
    Method & Mean RMSE & Std. Deviation & 95\% CI Lower & 95\% CI Upper \\
    \hline
    Proposed & 0.082 & 0.014 & 0.076 & 0.088 \\
    Classical PID                   & 0.215 & 0.042 & 0.197 & 0.233 \\
    Nominal SMPC                    & 0.131 & 0.027 & 0.119 & 0.143 \\
    Fixed FT-MPC                    & 0.126 & 0.024 & 0.115 & 0.137 \\
    NN-MPC                          & 0.183 & 0.039 & 0.166 & 0.200 \\
    \hline
    \end{tabular}
    \label{tab:stat_rmse}
    \end{table}
We validated our PIML-MPC framework through hardware experiments using the Crazyflie 2.1 nano-quadrotor, with an STM32F405RG (168 MHz), 192 KB RAM, and 1 MB Flash, Cycle counter (DWT) on STM32 to record per-iteration execution time (average over 1000 cycles). Custom C++ implementation (no OS, bare-metal) with optimized linear algebra (CMSIS-DSP library). The platform supports various sensor decks for autonomous operation (see manufacturer details\footnote{https://www.bitcraze.io/}). The training Dataset for Data-Driven Controllers (NN-MPC, Proposed) includes flight data from Crazyflie quadrotor (hover + trajectory tracking) with Dryden wind disturbance, with a dataset size of 10,000 samples (time-series, no overlap) for both NN-MPC and proposed PIML. Data preprocessing with Normalization to $[-1,1]$ using the mean/std of the training data (mean = $x_{\text{ref}}$, std = 0.1 for all states/inputs), and finally, the test dataset is 2,000 independent samples (unseen during training) to validate generalization. All MPC-based controllers (SMPC, FT-MPC, proposed method) use identical solver settings for fairness with CasADi 3.7.2 with IPOPT 3.14.17 (interior-point method) having parameters: $\text{tol} = 10^{-4}$, $\text{max\_iter} = 100$, $\text{print\_level} = 0$ (silent mode) with $N=15$ (fixed for all MPC variants) and Sampling time: $\Delta t = 0.01$ s (100 Hz control frequency, consistent with quadrotor hardware).
Table \ref{tab:comp_perf} summarizes average computation time and worst-case latency for all controllers (measured on STM32F405RG).
\begin{table}[htbp!]
\centering
\caption{Embedded Computational Performance (STM32F405RG)}
\label{tab:comp_perf}
\footnotesize
\begin{tabular}{lcc}
\hline
Controller & Avg. Computation Time (ms) & Worst-Case Latency (ms) \\
\hline
PID & 0.8 & 1.2 \\
SMPC & 8.4 & 10.1 \\
FT-MPC & 9.2 & 11.5 \\
NN-MPC & 15.6 & 18.3 \\
Proposed Method & 9.8 & 12.4 \\
\hline
\end{tabular}
\end{table}
\begin{figure*}[htbp!]
    \centering
    \includegraphics[width=1\linewidth]{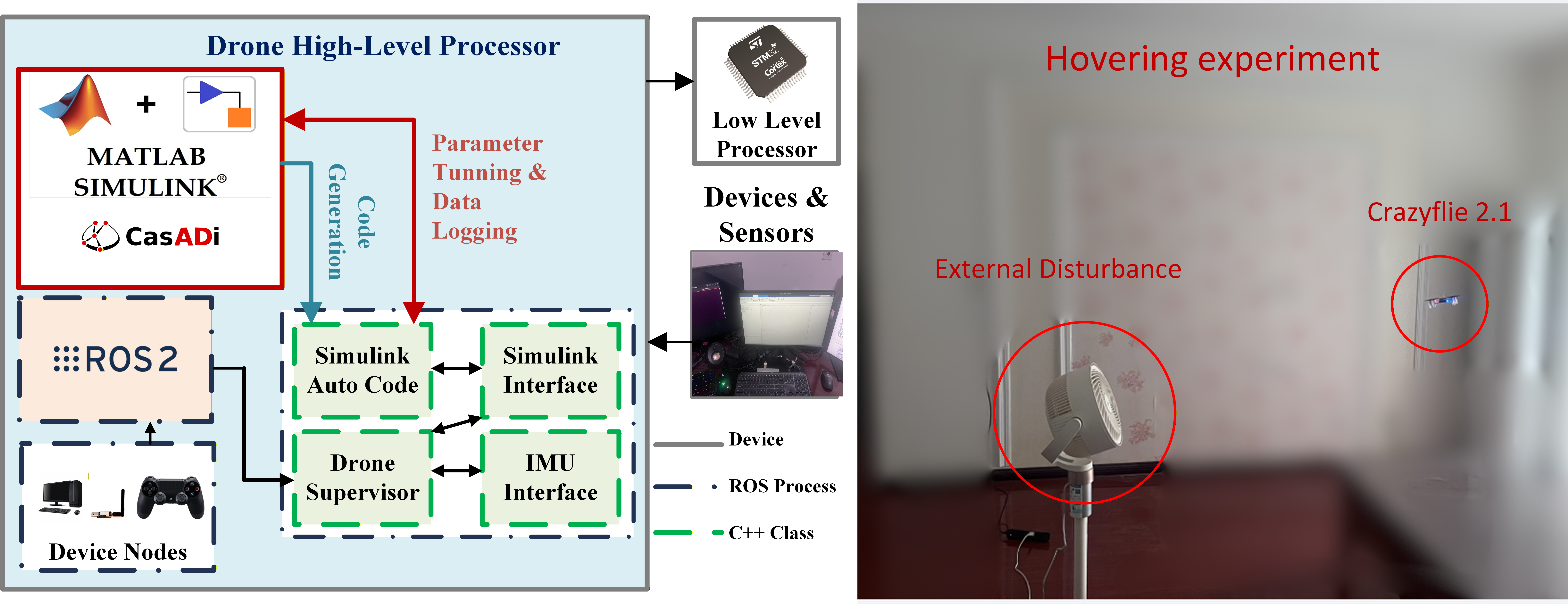}
    \caption{Simulation to hardware deployment and hovering experiments}
    \label{fig:7}
\end{figure*}
%
\begin{figure*}[ht!]
    \centering
    \includegraphics[width=.95\linewidth]{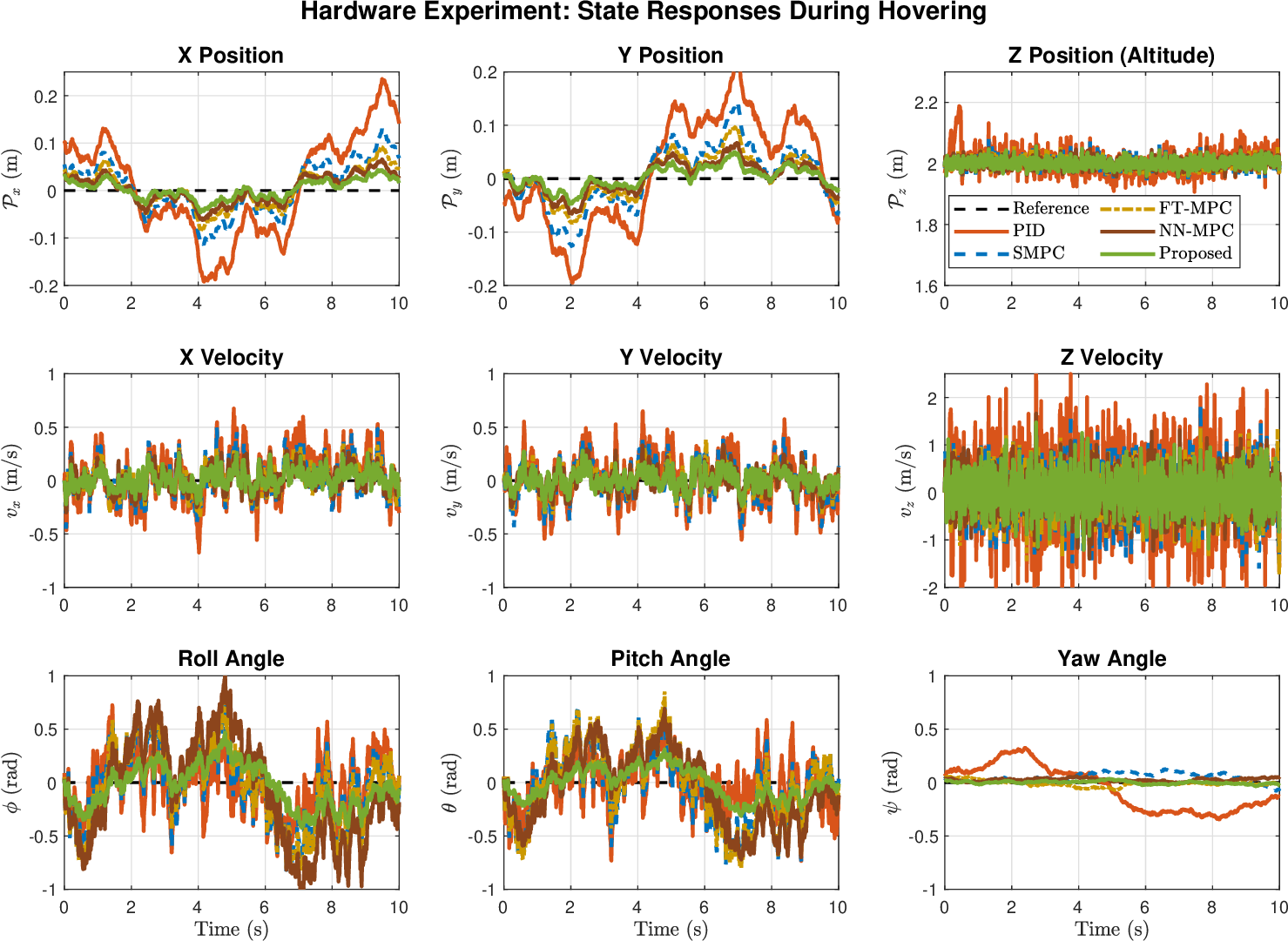}
    \caption{Real-time indoor experiments related to state response}
    \label{fig:8}
\end{figure*}
\begin{figure}[htbp!]
    \centering
    \includegraphics[width=.7\linewidth]{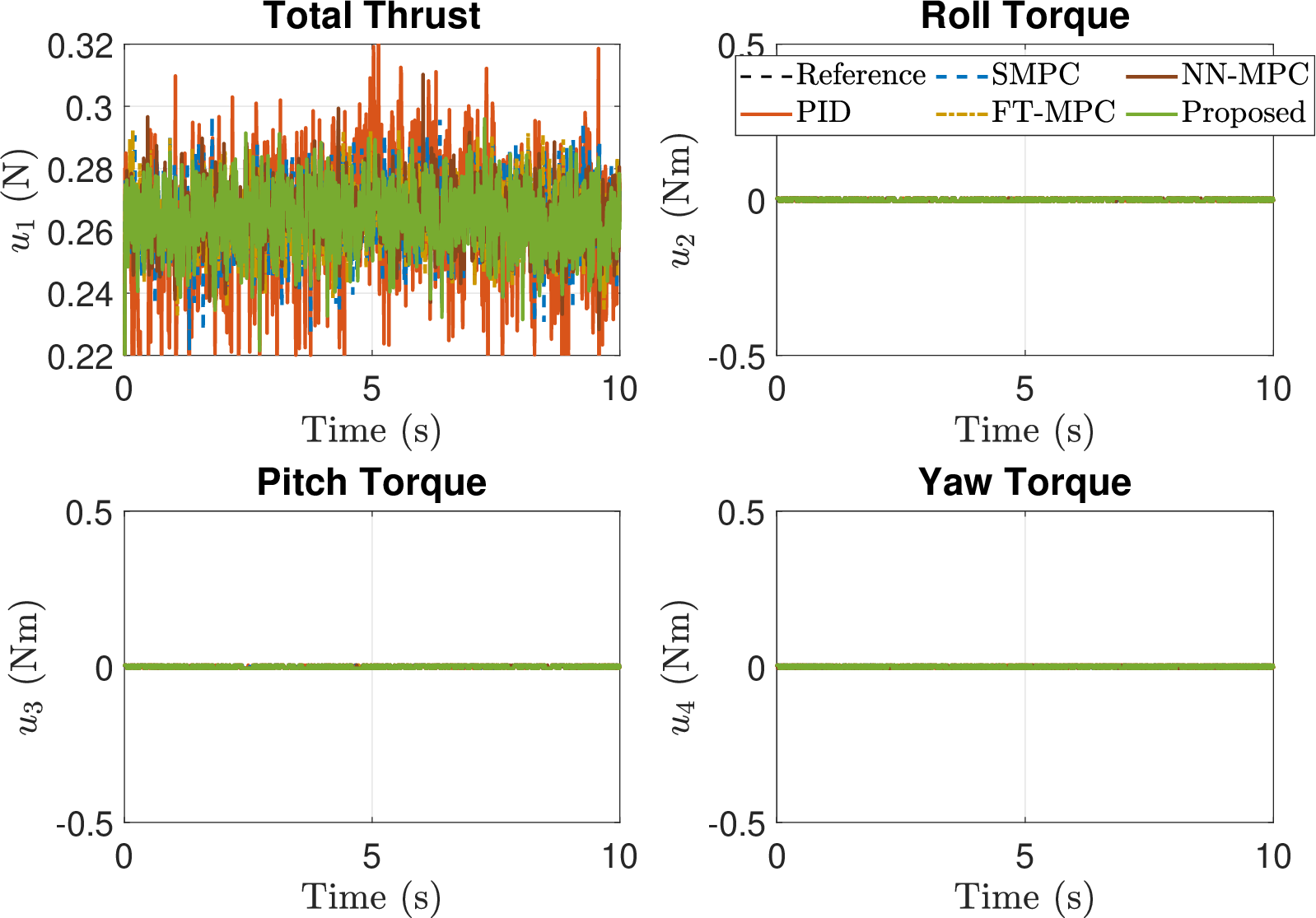}
    \caption{Control input during hovering experiment}
    \label{fig:9}
\end{figure}%
\begin{table*}[htbp!]
\centering
\caption{RMSE (m) and average computation time per iteration (ms) across different trajectories.}
\label{tab:tracking_errors}
\footnotesize
\begin{tabular}{l *{6}{S[table-format=1.4]}}
\toprule
\textbf{Trajectory type} & \textbf{PID} & \textbf{SMPC} & \textbf{FT-MPC} & \textbf{NN-MPC} & \textbf{Proposed} \\
\midrule
Helical (m)    & 0.3686 & 0.2516 & 0.1530 & 0.0520 & 0.0158 \\
Spline (m)    & 0.2000 & 0.1760 & 0.3524 & 0.1053 & 0.0183 \\
Lemniscate (m) & 0.1061 & 0.0701 & 0.0800 & 0.0303 & 0.0182 \\
\midrule
\textbf{Average RMSE (m)} & 0.2249 & 0.1659 & 0.1951 & 0.0625 & 0.0174 \\
\textbf{Average computation time (ms)} & 0.12 & 8.45 & 10.23 & 15.67 & 9.81 \\
\bottomrule
\end{tabular}
\end{table*}
Fig.~\ref{fig:7} illustrates the deployment from simulation to hardware, including hovering experiments, with real-time results shown in Fig.~\ref{fig:8}. Comparative analysis reveals that the presented method exhibits a relatively damped response with minimal oscillations compared to other approaches. Fig.~\ref{fig:9} shows control inputs, where our developed method produces smoother signals with reduced high-frequency content, indicating better energy efficiency and reduced motor wear. Quantitative metrics in Table~\ref{tab:tracking_errors_2} confirm that our approach outperforms alternatives across key performance indicators.
\begin{table}[htbp!]
\centering
\caption{Hardware experiment summary (RMSE).}
\label{tab:tracking_errors_2}
\footnotesize
\begin{tabular}{cccc}
\hline
\textbf{Method} & \textbf{Position (m)} & \textbf{Attitude (rad)} & \textbf{Max Altitude Error} (m) \\
\hline
PID        &  0.1552&              0.4413   &            0.1139 \\
SMPC       &   0.0883     &         0.4649     &         0.0763 \\
FT-MPC      &    0.0612   &           0.4743  &            0.0599  \\
NN-MPC      &    0.046   &           0.4621   &           0.0534  \\
Proposed   &       0.0326  &            0.2031  &             0.0506 \\
\hline
\end{tabular}
\end{table}
\subsection{Memory Consumption on STM32F405RG}
Memory usage is measured via STM32CubeIDE (linker script analysis + runtime RAM monitoring) and reported in Table \ref{tab:mem_perf} (critical for embedded systems with limited resources):
\begin{table}[h!]
\centering
\caption{Embedded Memory Consumption (STM32F405RG)}
\label{tab:mem_perf}
\footnotesize
\begin{tabular}{lcc}
\hline
Controller & Flash Usage (KB) & RAM Usage (KB) \\
\hline
PID & 8 & 2 \\
SMPC & 42 & 18 \\
FT-MPC & 56 & 24 \\
NN-MPC & 128 & 64 \\
Proposed Method & 66 & 32 \\
\hline
\end{tabular}
\end{table}
\paragraph{Memory Allocation Breakdown (Proposed Method)}: Static RAM (constants, lookup tables): 8 KB, Dynamic RAM (disturbance history: 50 samples × 12D × 4 bytes = 2.4 KB; RPI set: 12×12 matrix = 0.576 KB; library coefficients: 100 terms × 12D = 4.8 KB), MPC Solver RAM (CasADi workspace): 16.2 KB, Total RAM: 32 KB (16.7\% of STM32F405RG's 192 KB RAM), which is well within hardware limits.

The disturbance history (2.4 KB) contributes minimally to overall RAM usage, and the proposed method's memory footprint is 50\% lower than NN-MPC, validating its efficiency for resource-constrained embedded platforms.
\subsection{Discussion, limitations and Practical Insights}
The proposed PIML-MPC framework demonstrates several distinct advantages over existing control approaches. The enhanced tracking accuracy of the presented framework, evidenced by reduced RMSE error compared to both SMPC and PID controllers, is directly attributable to the learned residual dynamics model. This model captures aerodynamic and inertial effects not represented in the nominal first-principles model, improving prediction fidelity. The sparse structure of the ML-based model enables computationally efficient implementation. While the adaptive tube mechanism incurs additional computational load compared to non-adaptive approaches, the overall computational requirement remains feasible for real-time implementation on embedded hardware, as demonstrated in hardware experiments. The adaptive tube mechanism provides robustness against model uncertainty and disturbances while reducing conservatism relative to fixed-tube robust MPC formulations. This adaptation is informed by online estimates of the residual model error, allowing constraint satisfaction without excessively restricting the feasible control space. The data efficiency of the PIML approach enables effective model learning from limited flight data. The sparse regression framework, combined with physics-informed library functions, requires substantially fewer training samples than black-box NN-based alternatives to achieve comparable model accuracy. Compared with state-of-the-art learning-based MPC represented by NN-MPC, the proposed PIML sparse identification framework exhibits distinct advantages. First, benefiting from embedded physical prior knowledge and automatic sparse library construction, our method achieves satisfactory modeling accuracy with a limited number of flight samples, while black-box neural network MPC requires massive labeled datasets for training and suffers from poor generalization under unseen disturbances. Second, the learned sparse dynamics model has explicit analytical expressions that are fully interpretable, whereas NN models are black boxes and cannot provide physical explanations for system behavior. In terms of embedded deployment on resource-constrained quadrotors, the proposed method has a much smaller memory footprint (32 KB RAM) than NN-MPC (64 KB RAM), and its average computation time is reduced by 37.4\%. In addition, when combined with the real-time adaptive tube mechanism, our framework maintains robust constraint satisfaction under time-varying wind disturbances and model errors, outperforming conventional learning-based MPC without a dedicated robust design.

The proposed framework presents certain limitations that warrant consideration: (1) During the early learning phase (before sufficient flight data accumulates to refine the PIML model and RPI set), constraint tightening is intentionally conservative to ensure safety. This restricts maneuver aggressiveness to avoid constraint violations while the library and disturbance bounds adapt. The conservatism diminishes as more data is collected; (2) Discrete, periodic PIML model updates (implemented every $T_{\text{learn}}=5$ seconds in our experiments) may introduce minor, short-lived transients in the control signal. These transients arise from small adjustments to the nominal $\hat{f}^d$, and RPI set $\mathbb{S}(k)$ during updates, though they are bounded by the tube constraints and do not compromise stability. Continuous or event-triggered update strategies may mitigate this effect by aligning updates with meaningful model improvements rather than fixed time intervals; (3) The adaptive disturbance set update requires storing a short history of realized disturbances $\bm\Delta^{d}_{vw}(k)$ to compute $\lambda$ and $\gamma$. This increases memory overhead by ~8 KB (for 12-dimensional disturbances stored as 32-bit floats) relative to non-adaptive robust controllers (which use static disturbance bounds). While negligible for modern embedded MCUs (e.g., STM32F405RG with 192 KB RAM), it may be a consideration for ultra-low-power platforms with limited memory; (4) The efficacy of the sparse identification process is contingent upon the completeness of the candidate function library; an overly restrictive (incomplete) or redundant (bloated) library can degrade model accuracy. Critically, this does not imply manual library selection, as our library is constructed automatically via physics-required base terms and data-driven expansion. The limitation arises from the fixed degree of nonlinearity in auto-expanded terms-extreme dynamics (e.g., high-speed aerodynamic stall) may require higher-degree terms or specialized physics terms not included in the current auto-expansion rule.

Practical implementation yielded several insights. The adaptation gain parameter requires calibration based on the anticipated disturbance spectrum to balance convergence speed and noise sensitivity. $\lambda=0.9$ and $\gamma=0.95$ are empirically calibrated via repeated simulations and Crazyflie flight tests to reconcile noise suppression and dynamic adaptivity. Slight parameter changes within $\pm0.03$ yield tracking RMSE fluctuations below $3.5\%$, and the PIML sparse learning procedure stays unaffected since $\lambda$ and $\gamma$ merely regulate online disturbance-bound updating rather than the sparse regression for system identification. Closed-loop stability is preserved across such mild parameter perturbations. Note that the empirical selection of $\lambda$ and $\gamma$ does not constitute a critical drawback for the overall framework, as these hyperparameters only govern the recursive update of disturbance bounds instead of the core PIML identification and tube MPC formulation. Owing to the low performance sensitivity against mild parameter variations, the empirically chosen values ensure reliable real-world flight performance. Analytical optimal gain solving can be explored in future research to further remove empirical tuning. A safety fallback mode, triggered by metrics indicating model degradation, can enhance operational reliability. The control energy expenditure of the proposed method was observed to be intermediate between simpler PID controllers and more computationally intensive NN-based MPC. The computational complexity of the presented algorithm scales approximately linearly with the system state dimension, suggesting applicability to larger aerial platforms without prohibitive computational cost increase. These findings indicate that the developed framework provides a viable compromise between tracking precision, robustness guarantees, and computational demand, rendering it applicable to aerial vehicles operating under practical safety and resource constraints.
\section{Conclusion}\label{section5}
This paper presented a PIML-based MPC framework that leverages online PIML model learning with robust control guarantees. The key contributions include: (1) a PIML identification method that provides interpretable dynamics models, (2) an online disturbance set adaptation mechanism that reduces conservatism while maintaining robustness, and (3) formal guarantees of recursive feasibility and ISS. Numerical simulations and experiments demonstrated that our approach reduces computational load compared to SMPC, FT-MPC, and NN-MPC while maintaining comparable tracking performance. The adaptive tube mechanism outperformed FT-MPC by reducing conservatism without compromising safety.

Future work will focus on extending the framework to multi-vehicle systems and investigating distributed implementation strategies for swarm applications. Additionally, we plan to explore transfer learning techniques to reduce the initial data requirements for new vehicle configurations.
\appendix
\section{Proof of Lipschitz Continuity for the Quadratic Stage Cost (Assumption \ref{assumption7})}
\label{app:lipschitz_cost}
Assumption \ref{assumption7} states that the quadratic stage cost for the quadrotor MPC framework:
\[
\ell_c(\bm x,\bm u) = (\bm x - \bm x_{\text{ref}})^\top \bm Q (\bm x - \bm x_{\text{ref}}) + (\bm u - \bm u_{\text{ref}})^\top \bm R (\bm u - \bm u_{\text{ref}}),
\]
is Lipschitz continuous on the compact set $\mathbb{X} \times \mathbb{U}$. Here, the state set $\mathbb{X}\subset\mathbb{R}^{12}$ and input set $\mathbb{U} \subset \mathbb{R}^4$ are defined by quadrotor physical constraints, with $\bm Q \succeq 0$ (positive semi-definite) and $\bm R \succ 0$ (positive definite). We prove this property below.
\begin{proof}
Let $(\bm x_1, \bm u_1), (\bm x_2, \bm u_2) \in \mathbb{X} \times \mathbb{U}$ be arbitrary state-input pairs within the compact constraint set. Define:
\begin{align*}
\Delta \bm x &= \bm x_1 - \bm x_2, \quad \Delta \bm u = \bm u_1 - \bm u_2, \\
\bm x_0 &= \bm x_{\text{ref}}, \quad \bm u_0 = \bm u_{\text{ref}}.
\end{align*}

First, expand the difference of the stage cost:
\begin{align*}
    &\ell_c(\bm x_1, \bm u_1) - \ell_c(\bm x_2, \bm u_2) 
    = \underbrace{(\bm x_1 - \bm x_0)^\top \bm Q (\bm x_1 - \bm x_0) - (\bm x_2 - \bm x_0)^\top \bm Q (\bm x_2 - \bm x_0)}_{\text{State cost difference}} \\
    &+ \underbrace{(\bm u_1 - \bm u_0)^\top \bm R (\bm u_1 - \bm u_0) - (\bm u_2 - \bm u_0)^\top \bm R (\bm u_2 - \bm u_0)}_{\text{Input cost difference}}.
\end{align*}
Step 1: Simplify the quadratic form difference.\\
Use the identity for symmetric matrices ($M = M^\top$):
\[
\bm a^\top M \bm a - \bm b^\top M \bm b = (\bm a - \bm b)^\top M (\bm a + \bm b - 2\bm c),
\]
where $\bm c$ denotes a constant vector. For the state term, we set $\bm c = \bm x_0$, and for the input term $\bm c = \bm u_0$. Applying this identity:
\begin{align*}
(\bm x_1 - \bm x_0)^\top \bm Q (\bm x_1 - \bm x_0) - (\bm x_2 - \bm x_0)^\top \bm Q (\bm x_2 - \bm x_0) 
&= \Delta \bm x^\top \bm Q (\bm x_1 + \bm x_2 - 2\bm x_0), \\
(\bm u_1 - \bm u_0)^\top \bm R (\bm u_1 - \bm u_0) - (\bm u_2 - \bm u_0)^\top \bm R (\bm u_2 - \bm u_0) 
&= \Delta \bm u^\top \bm R (\bm u_1 + \bm u_2 - 2\bm u_0).
\end{align*}

Substitute back into the cost difference:
\[
\ell_c(\bm x_1, \bm u_1) - \ell_c(\bm x_2, \bm u_2) 
= \Delta \bm x^\top \bm Q (\bm x_1 + \bm x_2 - 2\bm x_0) + \Delta \bm u^\top \bm R (\bm u_1 + \bm u_2 - 2\bm u_0).
\]
Step 2: Apply the Cauchy-Schwarz Inequality.\\
For any vectors $\bm a, \bm b \in \mathbb{R}^n$, the Cauchy-Schwarz inequality reads:
\[
|\bm a^\top \bm b| \leq \|\bm a\| \cdot \|\bm b\|.
\]
Applying this to both terms:
\begin{align*}
\big|\Delta \bm x^\top \bm Q (\bm x_1 + \bm x_2 - 2\bm x_0)\big| 
&\leq \|\Delta \bm x\| \cdot \big\| \bm Q (\bm x_1 + \bm x_2 - 2\bm x_0) \big\|, \\
\big|\Delta \bm u^\top \bm R (\bm u_1 + \bm u_2 - 2\bm u_0)\big| 
&\leq \|\Delta \bm u\| \cdot \big\| \bm R (\bm u_1 + \bm u_2 - 2\bm u_0) \big\|.
\end{align*}

Step 3: Bounding terms via compactness.\\
Since $\mathbb{X} \times \mathbb{U}$ is closed and bounded (compact) due to quadrotor physical constraints, the terms $\bm x_1 + \bm x_2 - 2\bm x_0$ and $\bm u_1 + \bm u_2 - 2\bm u_0$ are uniformly bounded. Define:
\begin{align*}
\mathcal{L}_x &= \sup_{\bm x_1, \bm x_2 \in \mathbb{X}} \big\| \bm Q (\bm x_1 + \bm x_2 - 2\bm x_0) \big\|, \\
\mathcal{L}_u &= \sup_{\bm u_1, \bm u_2 \in \mathbb{U}} \big\| \bm R (\bm u_1 + \bm u_2 - 2\bm u_0) \big\|.
\end{align*}
Both $\mathcal{L}_x > 0$ and $\mathcal{L}_u > 0$ are finite constants, as the supremum of a continuous function over a compact set is bounded.

Step 4: Final Lipschitz bound.\\
Combining the above results yields the bound for the absolute cost difference:
\[
\big| \ell_c(\bm x_1, \bm u_1) - \ell_c(\bm x_2, \bm u_2) \big| 
\leq \mathcal{L}_x \cdot \|\Delta \bm x\| + \mathcal{L}_u \cdot \|\Delta \bm u\|.
\]
This satisfies the definition of Lipschitz continuity with Lipschitz constants $\mathcal{L}_x$ (for states) and $\mathcal{L}_u$ (for inputs). Therefore, the quadratic stage cost is Lipschitz continuous on $\mathbb{X} \times \mathbb{U}$.
\end{proof}
\section*{CRediT Authorship Contribution Statement}
\textbf{Tayyab Manzoor}: Writing – review \& editing, Project administration, Conceptualization, Methodology, Software, Validation, Formal analysis, Investigation, Data Curation, Visualization, Project administration, Funding acquisition. \textbf{Yasir Ali}: Writing – review \& editing, Visualization, Validation, Formal analysis, Funding acquisition. \textbf{Yuanqing Xia}: Writing – review \& editing, Supervision, Visualization, Funding acquisition, Project administration. \textbf{Lijie You}: Writing – review \& editing, Visualization, Validation, Project administration. \textbf{Yan Wang}: Writing – review \& editing, Visualization, Validation, Project administration.
\section*{Acknowledgment}
This work was supported in part by the Beijing Natural Science Foundation Haidian Original Innovation Joint Fund Project under Grant L252035, in part by the National Natural Science Foundation of China under Grant U25A20460, in part by the National Natural Science Foundation of China for International Young Scientists under Grant W2533176, in part by the Beijing Municipal Natural Science Foundation for Foreign Scholars under Grant IS25064, Henan Province Key International Science and Technology Cooperation Project under Grant 251111520400, in part by the Joint Fund Project of Shandong Provincial Natural Science Foundation under Grant ZR2025LZH001, in part by the Henan Provincial Talent Program under Grant 264000510006 and in part by the Henan Postdoctoral Foundation under Grant HN2026058.

\bibliography{mybib}

@ARTICLE{tayyab_tnnls1,
  author={Manzoor, Tayyab and Pei, Hailong and Xia, Yuanqing and Sun, Zhongqi and Ali, Yasir},
  journal={IEEE Transactions on Neural Networks and Learning Systems}, 
  title={Compound Learning-Based Model Predictive Control Approach for Ducted-Fan Aerial Vehicles}, 
  year={2025},
  volume={36},
  number={5},
  pages={9395-9407},
  doi={10.1109/TNNLS.2024.3422401}}

@ARTICLE{active_learning,
  author={Saviolo, Alessandro and Frey, Jonathan and Rathod, Abhishek and Diehl, Moritz and Loianno, Giuseppe},
  journal={IEEE Transactions on Robotics}, 
  title={Active Learning of Discrete-Time Dynamics for Uncertainty-Aware Model Predictive Control}, 
  year={2024},
  volume={40},
  number={},
  pages={1273-1291},
  doi={10.1109/TRO.2023.3339543}}

@ARTICLE{10606056,
  author={Tagliabue, Andrea and How, Jonathan P.},
  journal={IEEE Transactions on Robotics}, 
  title={Efficient Deep Learning of Robust Policies From {MPC} Using Imitation and Tube-Guided Data Augmentation}, 
  year={2024},
  volume={40},
  number={},
  pages={4301-4321},
  doi={10.1109/TRO.2024.3431988}}

@Article{drones7010004,
AUTHOR = {Manzoor, Tayyab and Pei, Hailong and Sun, Zhongqi and Cheng, Zihuan},
TITLE = {Model Predictive Control Technique for Ducted Fan Aerial Vehicles Using Physics-Informed Machine Learning},
JOURNAL = {Drones},
VOLUME = {7},
YEAR = {2023},
NUMBER = {1},
ARTICLE-NUMBER = {4},
ISSN = {2504-446X},
DOI = {10.3390/drones7010004}
}

@ARTICLE{TIE2,
  author={Li, Yuan and Yang, Xuebo and Zheng, Xiaolong},
  journal={IEEE Transactions on Industrial Electronics}, 
  title={Disturbance-Learning-Based Robust Model Predictive Control for Attitude Tracking of Small Aircraft}, 
  year={2025},
  volume={},
  number={},
  pages={1-11},
  doi={10.1109/TIE.2025.3536558}}

@Inbook{Sontag2008,
author="Sontag, Eduardo D.",
title="Input to State Stability: Basic Concepts and Results",
bookTitle="Nonlinear and Optimal Control Theory: Lectures given at the C.I.M.E. Summer School held in Cetraro, Italy June 19--29, 2004",
year="2008",
publisher="Springer Berlin Heidelberg",
address="Berlin, Heidelberg",
pages="163--220",
isbn="978-3-540-77653-6",
doi="10.1007/978-3-540-77653-6_3"
}

@Inbook{Limon2009,
author="Limon, D.
and Alamo, T.
and Raimondo, D. M.
and de la Pe{\~{n}}a, D. Mu{\~{n}}oz
and Bravo, J. M.
and Ferramosca, A.
and Camacho, E. F.",
title="Input-to-State Stability: A Unifying Framework for Robust Model Predictive Control",
bookTitle="Nonlinear Model Predictive Control: Towards New Challenging Applications",
year="2009",
publisher="Springer Berlin Heidelberg",
address="Berlin, Heidelberg",
pages="1--26",
isbn="978-3-642-01094-1",
doi="10.1007/978-3-642-01094-1_1"
}

@ARTICLE{NN-mpc1,
  author={Kong, Chuixu and Wang, Bin and He, Liang and Jiang, Qinglong and Liu, Chang},
  journal={IEEE Robotics and Automation Letters}, 
  title={{ANN-MPC}: An Adaptive Neural Network-Based Nonlinear Model Predictive Control Method for Robust Trajectory Tracking of Quadrotors}, 
  year={2025},
  volume={10},
  number={10},
  pages={10966-10973},
  doi={10.1109/LRA.2025.3607272}}

@ARTICLE{NN-MPC3,
  author={Kamath, Archit Krishna and Anavatti, Sreenatha G. and Feroskhan, Mir},
  journal={IEEE Transactions on Cybernetics}, 
  title={A Physics-Informed Neural Network Approach to Augmented Dynamics Visual Servoing of Multirotors}, 
  year={2024},
  volume={54},
  number={11},
  pages={6319-6332},
  doi={10.1109/TCYB.2024.3413072}}

@ARTICLE{rmpc1,
  author={Sun, Haidi and Dai, Li and Wang, Peizhan},
  journal={IEEE Transactions on Aerospace and Electronic Systems}, 
  title={An Efficient Moving Obstacle Avoidance Scheme for {UAV}s via Output Feedback Robust MPC}, 
  year={2024},
  volume={60},
  number={5},
  pages={6199-6212},
  doi={10.1109/TAES.2024.3401094}}

@INPROCEEDINGS{safe_learning,
  author={Drgoňa, Ján and Nghiem, Truong X. and Beckers, Thomas and Fazlyab, Mahyar and Mallada, Enrique and Jones, Colin and Vrabie, Draguna and Brunton, Steven L. and Findeisen, Rolf},
  booktitle={2025 American Control Conference (ACC)}, 
  title={Safe Physics-informed Machine Learning for Dynamics and Control}, 
  year={2025},
  volume={},
  number={},
  pages={591-606},
  doi={10.23919/ACC63710.2025.11107836}}

@Article{Andersson2019,
  author = {Joel A E Andersson and Joris Gillis and Greg Horn
            and James B Rawlings and Moritz Diehl},
  title = {{CasADi} -- {A} software framework for nonlinear optimization
           and optimal control},
  journal = {Mathematical Programming Computation},
  volume = {11},
  number = {1},
  pages = {1--36},
  year = {2019},
  publisher = {Springer},
  doi = {10.1007/s12532-018-0139-4}
}

@Article{watcher,
author={Andreas Wächter and Lorenz T. Biegler},
title={On the implementation of an interior-point filter line-search algorithm for large-scale nonlinear programming},
journal={Mathematical Programming},
volume={106},
pages={25-57},
year={2006},
doi={10.1007/s10107-004-0559-y},
publisher={Springer},

}

@ARTICLE{controlsystem_letter,
  author={Wang, Jiarui and Fazlyab, Mahyar},
  journal={IEEE Control Systems Letters}, 
  title={Actor–Critic Physics-Informed Neural Lyapunov Control}, 
  year={2024},
  volume={8},
  number={},
  pages={1751-1756},
  doi={10.1109/LCSYS.2024.3416235}}

@ARTICLE{tube_1,
  author={Qin, Wenchao and Tang, Wentao and Xia, Weiguo},
  journal={IEEE Transactions on Automation Science and Engineering}, 
  title={Robust Fault-Tolerant Control Based on Set-Membership Estimation and Tube {MPC}}, 
  year={2025},
  volume={22},
  number={},
  pages={19499-19510},
  doi={10.1109/TASE.2025.3594583}}

@ARTICLE{tube_2,
  author={Chen, Linbin and Liu, Yafei and Sun, Zhanbo and Yang, Lei and Yan, Qiruo},
  journal={IEEE Transactions on Vehicular Technology}, 
  title={A Tube-based MPC Method for Path Tracking of Autonomous Vehicles with Disturbances at Handling Limits}, 
  year={2025},
  volume={},
  number={},
  pages={1-15},
  doi={10.1109/TVT.2025.3620016}}

@INPROCEEDINGS{tube_3,
  author={Lin, Huifan and Zhang, Langwen and Xie, Wei},
  booktitle={2025 37th Chinese Control and Decision Conference (CCDC)}, 
  title={Trajectory Tracking Control of Quadrotor {UAV} with Tube-Based Model Predictive Control}, 
  year={2025},
  volume={},
  number={},
  pages={3197-3202},
  doi={10.1109/CCDC65474.2025.11090705}}

@ARTICLE{tube_4,
  author={Mishra, Prabhat K. and Gasparino, Mateus V. and Chowdhary, Girish},
  journal={IEEE Transactions on Automatic Control}, 
  title={Deep Model Predictive Control With Stability Guarantees}, 
  year={2025},
  volume={70},
  number={8},
  pages={5460-5467},
  doi={10.1109/TAC.2025.3550072}}

@ARTICLE{Robust_mpc2,
  author={Heinlein, Moritz and Subramanian, Sankaranarayanan and Lucia, Sergio},
  journal={IEEE Transactions on Automatic Control}, 
  title={Robust Model Predictive Control Exploiting Monotonicity Properties}, 
  year={2025},
  volume={70},
  number={9},
  pages={6260-6267},
  doi={10.1109/TAC.2025.3558137}}

@article{ARYAN2026114379,
title = {Physics-informed machine learning for precision Unmanned Aerial Vehicle control: Adaptive transformers with safety guarantees},
journal = {Engineering Applications of Artificial Intelligence},
volume = {172},
pages = {114379},
year = {2026},
issn = {0952-1976},
doi = {10.1016/j.engappai.2026.114379},
author = {Prakash Aryan and Sebastiano Panichella}
}

@Article{drones9030187,
AUTHOR = {Abdulkadirov, Ruslan and Lyakhov, Pavel and Butusov, Denis and Nagornov, Nikolay and Kalita, Diana},
TITLE = {Physics-Aware Machine Learning Approach for High-Precision Quadcopter Dynamics Modeling},
JOURNAL = {Drones},
VOLUME = {9},
YEAR = {2025},
NUMBER = {3},
ARTICLE-NUMBER = {187},
ISSN = {2504-446X},
DOI = {10.3390/drones9030187}
}

@article{OSMAN2026111646,
title = {Physics-informed sparse reinforcement learning for hybrid {VTOL UAV} control: {HILS} verification and tethered hover benchmarking},
journal = {Aerospace Science and Technology},
volume = {172},
pages = {111646},
year = {2026},
issn = {1270-9638},
doi = {10.1016/j.ast.2026.111646},
author = {Mohammed Osman and Yuanqing Xia and Mohammed Mahdi and Tayyab Manzoor and Ghulam E. Mustafa Abro and Abdulrahman H. Bajodah}
}

@INPROCEEDINGS{10770871,
  author={Osorio Quero, Carlos Alexander and Martinez-Carranza, Jose},
  booktitle={2024 21st International Conference on Electrical Engineering, Computing Science and Automatic Control (CCE)}, 
  title={Physics-Informed Machine Learning for {UAV} Control}, 
  year={2024},
  volume={},
  number={},
  pages={1-6},
  doi={10.1109/CCE62852.2024.10770871}}

@article{https://doi.org/10.1002/rnc.70272,
author = {Osman, Mohammed and Xia, Yuanqing and Mahdi, Mohammed and Manzoor, Tayyab and Bajodah, Abdulrahman H. and Ali, Asif and Ali, Abid and Ahmed, Azzam},
title = {An Adaptive {SIND}y-Lyapunov Model Predictive Control Framework for Dual-System {VTOL UAV}s},
journal = {International Journal of Robust and Nonlinear Control},
volume = {36},
number = {5},
pages = {2388-2417},
doi = {10.1002/rnc.70272},
year = {2026}
}

@article{hong2023physics,
  title={Physics-guided neural network and GPU-accelerated nonlinear model predictive control for quadcopter},
  author={Hong, Seong Hyeon and Ou, Junlin and Wang, Yi},
  journal={Neural Computing and Applications},
  volume={35},
  number={1},
  pages={393--413},
  year={2023},
  doi={https://doi.org/10.1007/s00521-022-07783-4},
  publisher={Springer}
}

@ARTICLE{9834096,
  author={Saviolo, Alessandro and Li, Guanrui and Loianno, Giuseppe},
  journal={IEEE Robotics and Automation Letters}, 
  title={Physics-Inspired Temporal Learning of Quadrotor Dynamics for Accurate Model Predictive Trajectory Tracking}, 
  year={2022},
  volume={7},
  number={4},
  pages={10256-10263},
  doi={10.1109/LRA.2022.3192609}}

@ARTICLE{11124830,
  author={Ghanifar, Mana and Ali Nikkhah, Amir and Kamzan, Milad and Teshnehlab, Mohammad and Tayefi, Morteza},
  journal={IEEE Access}, 
  title={Online Trajectory Regeneration for Multirotors via a Proportional-Derivative Physics-Informed Neural Network}, 
  year={2025},
  volume={13},
  number={},
  pages={145168-145190},
  doi={10.1109/ACCESS.2025.3598862}}

@article{YE2026111790,
title = {Safety-critical model predictive control for quadcopter {UAV} subject to wind disturbances and measurement errors in confined environments},
journal = {Aerospace Science and Technology},
volume = {173},
pages = {111790},
year = {2026},
issn = {1270-9638},
doi = {10.1016/j.ast.2026.111790},
author = {Hui Ye and Junjie Cao and Xiaofei Yang and Shuyi Shao}
}

@article{CHENG2026111513,
title = {Data-driven adaptive hybrid-based control for some tiltrotor {UAV}s in the low-speed flight},
journal = {Aerospace Science and Technology},
volume = {170},
pages = {111513},
year = {2026},
issn = {1270-9638},
doi = {10.1016/j.ast.2025.111513},
author = {Xiaomei Cheng and Yingxun Wang and Jiang Zhao and Ningjun Liu and Zhihao Cai}
}

@article{XU2026111789,
title = {Reinforcement learning-based distributed optimal formation tracking control with obstacle avoidance for quadrotor {UAV}s},
journal = {Aerospace Science and Technology},
volume = {173},
pages = {111789},
year = {2026},
issn = {1270-9638},
doi = {10.1016/j.ast.2026.111789},
author = {Lin-Xing Xu}
}

@article{TIAN2026111085,
title = {Adaptive model predictive control with extended state observer for unmanned aerial vehicle trajectory tracking},
journal = {Aerospace Science and Technology},
volume = {168},
pages = {111085},
year = {2026},
issn = {1270-9638},
doi = {10.1016/j.ast.2025.111085},
author = {Kun Tian and Lin-Jie Hu and Chao Bai and Hai-Peng Ren}
}

@article{IET_control,
author = {Lee, Jayden Dongwoo and Kim, Youngjae and Kim, Yoonseong and Bang, Hyochoong},
title = {Sparse Identification of Nonlinear Dynamics-Based Model Predictive Control for Multirotor Collision Avoidance},
journal = {IET Control Theory \& Applications},
volume = {19},
number = {1},
pages = {e70049},
doi = {https://doi.org/10.1049/cth2.70049},
year = {2025}
}

@INPROCEEDINGS{ICCTA,
  author={Rashad, Abdullah and Ghorab, Hashem and Othman, Mostafa and El-Badawy, Ayman},
  booktitle={2024 34th International Conference on Computer Theory and Applications (ICCTA)}, 
  title={Data-Driven {MPC} for Quadrotor {UAV}s Using {SIND}y}, 
  year={2024},
  volume={},
  number={},
  pages={165-170},
  doi={10.1109/ICCTA64612.2024.10974872}}
\bibliographystyle{elsarticle-num}
\end{document}